\documentclass[a4paper, amsfonts, amssymb, amsmath, reprint, showkeys, nofootinbib, twoside]{revtex4-2}
\usepackage[english]{babel}
\usepackage[utf8]{inputenc}
\usepackage[colorinlistoftodos, color=green!40, prependcaption]{todonotes}
\usepackage{amsthm}
\usepackage{mathtools}
\usepackage{physics}
\usepackage{xcolor}
\usepackage{graphicx}
\usepackage[left=23mm,right=13mm,top=35mm,columnsep=15pt]{geometry} 
\usepackage{adjustbox}
\usepackage{placeins}
\usepackage[T1]{fontenc}
\usepackage{lipsum}
\usepackage{csquotes}

\usepackage{amsmath}
\usepackage{bm}

\newcommand{\R}{\mathcal{R}}
\newcommand{\D}{\mathrm{d}}

\usepackage{xcolor}
\usepackage[pdftex, pdftitle={Article}, pdfauthor={Author}]{hyperref} 
\definecolor{snblue}{RGB}{48,65,121}
\definecolor{cinnamon}{RGB}{165,74,54}
\definecolor{sgreen}{RGB}{194,158,62}

\begin{document}
\title{Shaping dynamical neural computations using spatiotemporal constraints}

\author{Jason Z. Kim}
    \email[Correspondence email address: ]{jk2557@cornell.edu}
    \affiliation{\scriptsize Department of Physics, Cornell University, Ithaca, NY 14853, USA}
\author{Bart Larsen}
    \affiliation{\scriptsize Department of Pediatrics, Masonic Institute for the Developing Brain, University of Minnesota}
\author{Linden Parkes}
    \email[Correspondence email address: ]{linden.parkes@rutgers.edu}
    \affiliation{\scriptsize Department of Psychiatry, Rutgers University, Piscataway, NJ 08854, USA}

\date{\today} 

\begin{abstract}
Dynamics play a critical role in computation. The principled evolution of states over time enables both biological and artificial networks to represent and integrate information to make decisions. In the past few decades, significant multidisciplinary progress has been made in bridging the gap between how we understand biological \textit{versus} artificial computation, including how insights gained from one can translate to the other. Research has revealed that neurobiology is a key determinant of brain network architecture, which gives rise to spatiotemporally constrained patterns of activity that underlie computation. Here, we discuss how neural systems use dynamics for computation, and claim that the biological constraints that shape brain networks may be leveraged to improve the implementation of artificial neural networks. To formalize this discussion, we consider a natural artificial analog of the brain that has been used extensively to model neural computation: the recurrent neural network (RNN). In both the brain and the RNN, we emphasize the common computational substrate atop which dynamics occur---the connectivity between neurons---and we explore the unique computational advantages offered by biophysical constraints such as resource efficiency, spatial embedding, and neurodevelopment.
\end{abstract}


\maketitle
\newpage

\section{Introduction}
Dynamics have long underpinned computation. From the cycles of central pattern generators that support locomotion \cite{ijspeert2008central} to the networks of large-scale brain dynamics thought to regulate decision-making \cite{seeley2007dissociable}, it is clear that biological systems make ample use of their time-evolution to respond to their environment. Harnessing this dynamical computation, artificial recurrent neural networks (RNNs) have been trained to successfully perform the same computational tasks as humans \cite{mante2013context,werbos1990backpropagation}. However, while inspired by the brain, training of RNNs is typically carried out in an unconstrained manner, leading to solutions that lack biophysical realism. Additionally, decades of neuroscience research has demonstrated the importance of biological constraints for achieving the brain's unique structure and capabilities \cite{betzel_generative_2016, oldham_development_2019, oldham_modeling_2022, akarca_generative_2021}. Here, we draw on literature from dynamical systems and neuroscience to discuss (i) how RNNs leverage dynamics to compute, and (ii) how biophysical constraints may shape this computation by guiding the formation of network structure. At the intersection of these goals exists an opportunity to study how biologically-constrained RNNs may yield more powerful and more interpretable computational models.
~\\

To understand how biologically realistic neurons compute, there has been a long and rich history of modeling and interpreting neurobiological systems to leverage their computational capabilities. These quantitative models fall under the category of \textit{dynamical systems}, whose evolution in time is determined by mathematical functions. At the scale of a single neuron, detailed circuit models of the ion channels that mediate membrane voltage have enabled quantitative understanding of signal propagation and computation in dendrites \cite{hodgkin1952quantitative,petousakis2022impact}. At the scale of neural populations, mean-field models of excitatory and inhibitory neurons have enabled the study of neural circuits for biological sensing, imitation, and attention \cite{wilson1972excitatory,pinto1996quantitative,sadeghi2020dynamic}. At the whole-brain level, both linear \cite{parkes_using_2023,lynn_physics_2019, seguin_brain_2023, srivastava_models_2020} and non-linear \cite{breakspear_dynamic_2017, roberts_metastable_2019} dynamical models have been used to simulate large-scale activity patterns, and have examined how those patterns spread across the brain's white matter tracts. Across this broad range of systems, scales, and models, there exists a diversity of ways in which dynamics can be used for computation, as well as a crucial dependence of these dynamics on biophysical parameters. 
~\\

More recently, with technological advances in deep learning, the study of neural computation has adopted a more functional direction that moves away from biological realism. That is, rather than seeking a direct biophysical model \cite{hodgkin1952quantitative,fitzhugh1961impulses,wilson1972excitatory,mcculloch1943logical}, RNNs posit a general dynamical system that is Turing complete \cite{chung2021turing}, with parameters that are trained to solve computational tasks. We focus on RNNs because they have been used extensively to understand how general brain-like systems leverage dynamics to perform computation. Examples of this use include time-series prediction \cite{liu2020dstp}, source-separation \cite{lu2020supervised}, decision-making \cite{mante2013context}, odor classification \cite{wang_evolving_2021}, as well as concurrent performance of multiple cognitive tasks \cite{yang_task_2019}. However, these insights and models often fail to translate into the real computational substrate of the brain---the neural architecture---because RNNs are trained without regard for biophysical constraints. 
~\\

To merge biological realism with computational dynamics, we must first understand the physical embedding and constraints of the brain. In contrast to artificial RNNs, the brain is embedded within a circumscribed physical space \cite{roberts_contribution_2016}, and its inter-connectivity is subject to limited metabolic resources \cite{bullmore_economy_2012}. This discrepancy makes it challenging to translate insights relating the structure, dynamics, and computation of biological brains to artificial RNNs. Neuroscience has studied these resource constraints for more than a century \cite{ramon_y_cajal_histology_1995}, suggesting that the brain is pressured to make efficient use of space, material, and time. That is, the brain must learn to communicate efficiently (time) while leveraging limited physical (space) as well as metabolic and cellular (material) resources. Critically, many of the brain's topological features of connectivity and communication are thought to emerge as a consequence of navigating these pressures \cite{bullmore_economy_2012,betzel_generative_2016, oldham_development_2019, oldham_modeling_2022, akarca_generative_2021}. These findings suggest that the brain's resource constraints play a critical role in shaping its dynamic repertoire and computational capacity. 
~\\

 Here, our goal is to lay out promising new directions for improving the computational power and interpretability of RNN models of the brain. We posit that this goal will be achieved by placing biological constraints on RNNs that shape their structure and activity in systematic ways, which will in turn produce computationally improved dynamics. We focus on two aspects of RNN computational dynamics: the diversity of information that is represented by the neurons (expressivity), and the manipulation of low-dimensional internal representations (latent-spaces). In each section, we examine how biology shapes brain networks---with particular emphasis on the spatially-patterned macro-scale organizing principles of the cortex---and discuss how these constraints may be ported to RNNs to improve performance with interpretable structure and dynamics. Overall, we discuss how insights from biological and artificial computation can enrich each other towards a new generation of biophysically realistic RNNs.

\section{The RNN Model}

\begin{figure}[!ht]
\begin{center}
\includegraphics[width=\linewidth]{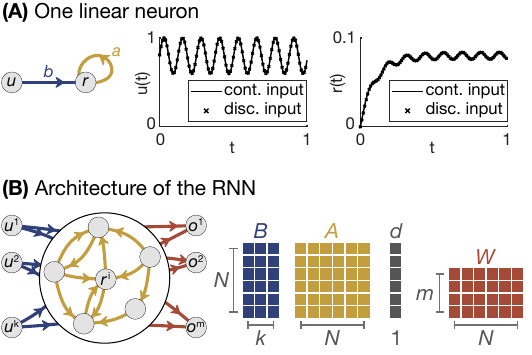}
\end{center}
\caption{\textbf{One-dimensional dynamics and RNN architecture.} {\bf{A}}, Schematic of a single neuron model of a leaky integrator with an input connection in blue, and a recurrent connection in gold (left), alongside examples of both a continuous and discretized input into the neuron (center), and the state of the input-driven neuron from the continuous (Eq.~\ref{eq:integrator_continuous}) and discretized (Eq.~\ref{eq:integrator_discrete}) dynamics (right). {\bf{B}}, Schematic of an RNN with $k$ inputs and $m$ outputs, with input connections and matrix in blue, recurrent connections and matrix in gold, and output connections and matrix in maroon.}
\label{fig:figure1_rnn}
\end{figure}

To mathematically model the time-evolution of neural systems, we turn to dynamical systems which posit that the next state of a neural system can be written as a function of the current state and an input as
\begin{align}
\label{eq:dynamical_discrete}
\bm{r}_{t+1} = f(\bm{r}_t,\bm{u}_t).
\end{align}
Here, $\bm{r}_t \in \R^n$ is a vector of $n$ neural activity states, $\bm{u}_t \in \R^k$ is a vector of $k$ inputs, and $f$ is a function. As an example, let us consider a simple leaky integrator model with a single neuron which evolves according to
\begin{align}
\label{eq:integrator_discrete}
r_{t+1} = ar_t + bu_t,
\end{align}
where $0 \leq a < 1$ and $b$ are real numbers (Fig.~\ref{fig:figure1_rnn}A). As time evolves forward, the neuron state integrates the input $bu_t$, and the accumulated history of inputs decays at a rate set by $a$.
~\\

While Eq.~\ref{eq:dynamical_discrete} is written with $t$ advancing in integer steps---thereby called a discrete-time dynamical system---many physical neural models evolve forward continuously in time as,
\begin{align}
\label{eq:dynamical_continuous}
\frac{\D}{\D t} \bm{r}(t) = \hat{f}(\bm{r}(t),\bm{u}(t)).
\end{align}
We can approximate these continuous dynamics as discrete by evolving Eq.~\ref{eq:dynamical_continuous} in time using steps of $\Delta t$ as,
\begin{align}
\label{eq:discretize}
\bm{r}_{t+\Delta t} = f(\bm{r}_t,\bm{u}_t) = \bm{r}_t + \int_{t}^{t+\Delta t} \hat{f}(\bm{r}(\tau),\bm{u}(\tau)) \D \tau.
\end{align}
For example, the continuous-time version of the leaky integrator neuron is given by
\begin{align}
\label{eq:integrator_continuous}
\frac{\mathrm{d}}{\mathrm{d}t} r(t) = \hat{a}r(t) + \hat{b}u(t),
\end{align}
for $\hat{a}\leq0$ (Fig.~\ref{fig:figure1_rnn}A). In this case, the parameters of these two models can be interchanged through the transformation $a = e^{\hat{a}\Delta t}$, $b = (a-1)\hat{b}/\hat{a}$, but fundamental differences exist between continuous- and discrete-time systems \cite{dupont2019augmented}. Regardless of the system type, neural models make tradeoffs between complexity in the level of detail and tractability.
~\\

The RNN is a model that attempts to capture the biophysical quantity of the \textit{interactions} between neurons through the connectivity matrix $A$. In tandem, the RNN simplifies the precise functional form of that interaction through the activation function $f$. In its most basic form, an RNN is a subset of dynamical systems (Eq.~\ref{eq:dynamical_continuous}) that evolves in time as
\begin{align}
\label{eq:RNN}
\begin{split}
\bm{r}_{t+1} &= f(\textcolor{sgreen}{A}\bm{r}_t + \textcolor{snblue}{B}\bm{u}_t + \bm{d}),\\
\bm{o}_t &= g(\bm{r}_t).
\end{split}
\end{align}
where $A \in \R^{n\times n}$ is the connectivity matrix between neurons, $B \in \R^{n\times k}$ is a matrix that linearly maps the inputs to the neurons, $\bm{d} \in \R^n$ is a vector of bias terms, and $f$ is an activation function (Fig.~\ref{fig:figure1_rnn}B). Rather than having $f$ be a complex and biophysically motivated function, it is often approximated as a simple nonlinear function such as a sigmoid. The output of the RNN, $\bm{o}_t$, is usually taken to be some function $g$ of the RNN state, and is often a linear output $\bm{o}_t = \textcolor{cinnamon}{W}\bm{r}_t$. Typically, $A,B,\bm{d}$, and $W$ are treated as learnable parameters, some or all of which can be trained using a wide variety of methods \cite{werbos1990backpropagation,lukovsevivcius2009reservoir,ali2022predictive}.
~\\

We focus primarily on the computational role of the connectivity matrix $A$, as it dictates how the information in the RNN states is integrated as in the leaky integrator example (Eq.~\ref{eq:integrator_discrete}, \ref{eq:integrator_continuous}). Despite its crucial importance in implementing computation, most uses of RNNs do not consider the biological pressures experienced by the brain while training RNN connectivity. In the following section, we describe how diversely the RNN states can \textit{express} the inputs by leveraging $A$, and how biological processes and constraints reflect, mediate and accentuate this diversity.

\section{Tuning Expressivity through Regularilized Activity and Neuromodulation} \label{expressivity}

When solving any computational problem, the expressivity of the language used is of crucial importance. The more expressive a language is, the greater the set of computations, formulae, and theorems comprise that language \cite{felleisen1991expressive,sipser1996introduction}. For example, a programming language that supports conditional \verb|if| statements can represent many more programs than an equivalent language that is without an \verb|if| statement.
~\\

In the same way, neural networks can be viewed from the lens of expressivity. Specifically, we can ask: given arbitrary weights, what is the space of functions or dynamics that can be achieved? Previous work has demonstrated that shallow multi-layer perceptrons (MLPs) are universal function approximators \cite{hornik1989multilayer,cybenko1989approximation}, and that RNNs are universal dynamics approximators \cite{schafer2006recurrent}. However, even if a particular function or dynamical pattern is theoretically achievable, artificial neural networks---just like biological neural networks---must be trained from an initial condition. Hence, the study of expressivity extends beyond theoretical guarantees, and has been shown to rely heavily upon architectural features such as depth \cite{poole2016exponential}, neuron activation \cite{raghu2017expressive}, and connectivity \cite{bertschinger2004real}, which are crucial for explaining and engineering the success of modern-day neural networks.
~\\

In this section, we study three consequences of biological processes for expressivity. First, we introduce RNN expressivity as a richness of time history information about the inputs, and tie this richness to spatially-patterned temporal receptive fields in the brain that underpin information integration. Second, we study constraints on expressivity induced by resource constraints on neural activity as a putative learning mechanism. Finally, we explore the potential for neural networks to modulate their expressivity at short time-scales through neuromodulation. Together, the processes of the brain offer enticing and novel paradigms for training and constructing more expressive RNNs under biological constraints. 

\subsection{Expressivity as variable time-lagged integration of information}

\begin{figure*}[!ht]
\begin{center}
\includegraphics[width=\textwidth]{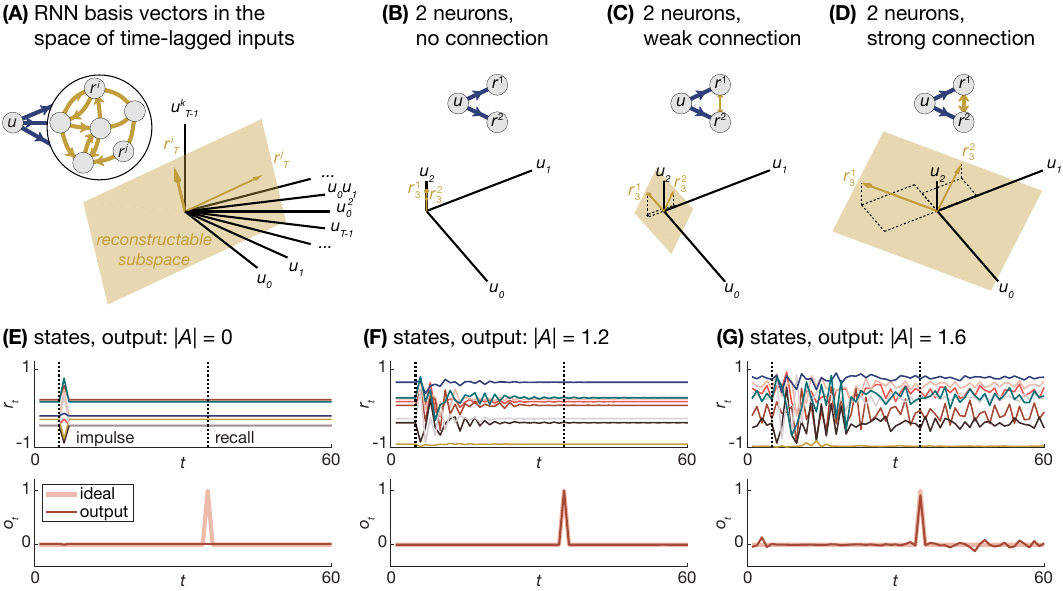}
\end{center}
\caption{\textbf{Expressivity and the representational basis formed by RNNs.} {\textbf{A}} Schematic of an RNN with a single input, the basis vectors formed by neurons $i$ and $j$, and the reconstructable subspace spanned by those vectors. {\textbf{B}} Schematic and reconstructable subspaces for a simple 2 neuron system with no recurrent connections, {\textbf{C}} weak recurrent connections, and {\textbf{D}} strong recurrent connections. {\textbf{E}} Example trajectories from a subset of 8 neurons of a 50 neuron system after receiving an impulse (top), alongside the ideal and true output of the RNN (bottom) after training a linear readout, $o_t = W\bm{r}_t$, to reproduce a time-lagged version of the impulse with $|A|=0$, {\textbf{F}} $|A|=1.2$, and {\textbf{G}} $|A|=1.6$.}
\label{fig:figure2_expressivity}
\end{figure*}

Expressivity of RNNs is intricately tied to the concept of \textit{stability}: how quickly a perturbation to the RNN state decays \cite{sompolinsky1988chaos}. If a perturbation decays very quickly, then the information contained in the perturbation cannot be used by the RNN for extended information processing. On the other hand, if the perturbation grows uncontrollably, then the temporal information in the inputs quickly becomes too complex to be represented by the finite number of neurons. As a result, an optimal amount of controlled stability should maximally preserve temporal information without saturating the RNN's capacity.
~\\

We can quantify this intuition through a simple recursive substitution of Eq.~\ref{eq:RNN},
\begin{align}
\label{eq:recursion}
\begin{split}
\bm{r}_{t+1} &= f(A\bm{r}_t + B\bm{u}_t + \bm{d})\\
&= f(A(f(A\bm{r}_{t-1} + B\bm{u}_{t-1} + \bm{d}) + B\bm{u}_t + \bm{d})\\
&= \dotsm\\
&= h(\bm{r}_0,\bm{u}_0,\bm{u}_1,\dotsm,\bm{u}_t).
\end{split}
\end{align}
Hence, in a noiseless system, the RNN state $\bm{r}_{t+1}$ can be written as an explicit function $h$ of the initial state, $\bm{r}_0$, and the full time history of the inputs, $\bm{u}_{\tau}$, mediated by the recursive application of $A, B, \bm{d}$, and the activation function $f$. The state of all neurons $\bm{r}_{t+1}$ generates a \textit{basis} for a subspace of the delay-embedded space of inputs $\bm{u}_{\tau}$ (Fig.~\ref{fig:figure2_expressivity}A), which means that the neuron states at time $t+1$ hold information about the time history of the inputs mediated by $A$, $B$, and $\bm{d}$ \cite{kim2023neural}. Thus, when computing an \textit{output} using the neural states, we are implicitly computing an output using a basis of time-lagged input terms, where the connectivity defines the basis vectors, and therefore the reconstructable subspace (Fig.~\ref{fig:figure2_expressivity}A). The more expressive this time-lagged basis, the greater the diversity of output functions which can be computed.
~\\

This expressivity is intimately tied to the connectivity matrix \cite{rajan2006eigenvalue}, and has been studied through many different lenses such as computation at the edge of chaos \cite{bertschinger2004real,langton1990computation,sussillo2009generating}, criticality, and avalanches \cite{munoz2018colloquium,hochstetter2021avalanches}. The RNN's stability is set by the specific activation function $f$ and the connectivity matrix $A$. To gain intuition for this dependence, let us consider a simple linear 2 neuron system driven by one input. When there is no connectivity between the neurons, they store no time history of the inputs, and their state at time $t=3$ is purely a function of the input at time $t=2$ (Fig.~\ref{fig:figure2_expressivity}B). When we add weak connections between the neurons, they begin to store some information about the input at the previous time point $t=1$ (Fig.~\ref{fig:figure2_expressivity}C). When we strengthen these connections, the neurons begin to store information from further back in time at $t=0$ (Fig.~\ref{fig:figure2_expressivity}D). 
~\\

To further develop this intuition for larger systems and a specific task, we consider a 50 neuron system whose connectivity is randomly initialized, and whose output is trained to recall an impulse from 30 time steps in the past. At the trivial limit of an RNN with no connectivity where $|A| = 0$, the recursion of Eq.~\ref{eq:recursion} yields $\bm{r}_{t+1} = f(B\bm{u}_t + \bm{d})$, and we see that there is no time history of the input present in the RNN state. As a result, the RNN is unable to recall the input at a later point in time (Fig.~\ref{fig:figure2_expressivity}E) \cite{ju2020network}. As we increase the strength of connectivity, the RNN state stores more information about longer time lags of the input, $\bm{u}_{t-\tau}$, and is thus able to more accurately recall the input later in time (Fig.~\ref{fig:figure2_expressivity}F). As the connectivity strength continues to further increase, the RNN state holds increasingly more time lags of the input until it saturates such that the number of neurons is smaller than the dimension of the space of time-lagged input functions, thereby forming an incomplete basis for that space (Fig.~\ref{fig:figure2_expressivity}G).
~\\

This link between storing long time histories and expressivity is displayed prominantly and spatially in the brain. Specifically, there exists a tight coupling between longer periods of temporal integration and higher-order computation, and this relationship varies systematically across the cortex \cite{wolff_intrinsic_2022}. At the macro-scale, cortical brain regions are thought to follow a dominant axis of variation that encodes a global processing hierarchy \cite{mesulam_representation_2008}. This gradient of brain organization is broadly referred to as the sensorimotor-association (S-A) axis \cite{sydnor_neurodevelopment_2021, margulies_situating_2016}. The S–A axis spans from primary cortices supporting sensation and movement at the bottom, to multimodal cortices supporting multisensory processing and integration in the middle, to transmodal association cortices supporting higher-order cognition at the top. The S-A axis is observed across multiple diverse features of brain structure and function \cite{sydnor_neurodevelopment_2021} and is conserved across species \cite{margulies_situating_2016,xu_cross-species_2020}, demonstrating its evolutionary roots. Notably, as regions traverse up the S-A axis they undergo a progressive lengthening of their temporal receptive windows \cite{hasson_hierarchy_2008, wolff_intrinsic_2022}. Specifically, regions' intrinsic functional timescales vary over the S-A axis \cite{gao_neuronal_2020, sydnor_intrinsic_2023} with regions at the top showing slower fluctuations reflecting longer temporal receptive windows. In turn, these longer windows are thought to enable greater accumulation and integration of information over time, facilitating higher-order cognition \cite{wolff_intrinsic_2022}. Conversely, regions at the bottom of the S-A axis show relatively fast dynamics, which is thought to underpin rapid integration of recent sensory information \cite{wolff_intrinsic_2022}.
~\\

This spatial patterning of receptive windows suggests that the brain---unlike naively constructed RNNs---distributes its computational expressivity systematically across the cortex, and research suggests that this may be critical for functional integration \cite{wolff_intrinsic_2022}. The above data suggest that RNNs too may benefit from spatially varying periods of temporal integration. Specifically, the stability of RNNs is typically only considered globally across the entire system, as the connectivity of many RNNs are initialized randomly. If the RNN is linear (i.e., if the activation function $f$ is the identity matrix), then any ensuing dynamic instabilities are localized to linear subspaces, or \textit{modes}, of neural activity \cite{hespanha2018linear}. However, if the RNN is nonlinear (e.g., $f = \tanh$), then instabilities bleed into other modes, making it difficult for the RNN to form a clean segregation of time-scales. Hence, varying periods of temporal integration could be achieved by specifying spatially varying penalties on neuronal timescales into the RNN cost functions. Such penalties may give rise to spatially segregated modules responsible for processing inputs at different time-scales. 

\subsection{Learning by suppressing activity}

While expressivity hinges on a careful balance of stable dynamics, biological neural networks are constrained by energy; more active neurons require more metabolic energy, which is a limited resource. As such, while artificial networks can maximize their expressivity through unconstrained backpropagation, the brain's capacity to learn is restricted by resource constraints; also, the extent to which the brain performs backpropagation remains unclear \cite{marblestone_toward_2016}, which has motivated the machine learning community to consider more biologically-inspired optimization approaches. Hence, penalizing activity in RNNs should intuitively penalize computational capability through reduced expressivity. However, recent work has instead demonstrated important computational benefits of minimizing energy usage during training \cite{ali_predictive_2022}. For example, Ali \emph{et al.} \cite{ali_predictive_2022} trained an RNN to predict sequences of handwritten digits, and examined how different optimization functions impacted model architecture and behavior. Specifically, Ali \emph{et al.} \cite{ali_predictive_2022} did not train their RNN to minimize prediction error through backpropagation. Instead, they trained their model to minimize absolute levels of neural activity prior to passing that activity through neurons' activation functions (here, ReLU). Such \emph{preactivation minimization} is akin to selectively minimizing neurons' presynaptic inputs in biological networks. Critically, executing this cost function required no information about task performance, and instead simply limited the RNN's resources in a biologically plausible way. 
~\\

Alongside good task performance, Ali \emph{et al.} \cite{ali_predictive_2022} observed dynamics in their RNN indicative of predictive coding. Predictive coding describes the hypothesis that the brain stores and updates expectations about its environment which it compares with incoming sensory evidence for those expectations \cite{millidge_predictive_2021}. Specifically, Ali \emph{et al.} \cite{ali_predictive_2022} observed activity patterns in their RNN suggestive of (i) selective self-inhibition in neurons receiving visual stimuli and (ii) prediction of future inputs in neurons not receiving visual stimuli. These results accord with hierarchically-organized predictive coding coupled to the S-A axis, wherein association cortices store predictions and via their distributed connectivity modulate activity in sensorimotor cortices \cite{bastos_canonical_2012, singer_recurrent_2021, friston_computational_2022}. Taken together, the findings of Ali \emph{et al.} \cite{ali_predictive_2022} indicate that while limiting the neurons' activation might intuitively reduce their expressivity---thereby limiting their computational capability in RNNs---it can also serve as a unique mechanism for distributed learning that is more biophysically realistic than backpropagation.

\subsection{Dynamically Tuning Expressivity \textit{via} Neuromodulation}

The preceding sections discussed expressivity as an emergent property of a trained network that can be modified by placing certain constraints on RNN training. This static expressivity has been shown to be effective at performing a wide range of tasks, and is the basis of the success of \textit{reservoir computing} \cite{jaeger2002tutorial}. Unlike in an RNN, the internal connectivity of the reservoir computer (RC) is not trained. Instead, only the output is trained, typically as a weighted sum of RC states  \cite{lukovsevivcius2009reservoir}. Hence, RCs rely completely on the preexisting expressivity of their internal dynamics to generate a sufficiently expressive basis representation of their inputs. Because RCs can be trained without knowledge or modification of the internal system, a wide variety of physical systems have been explored as efficient RCs \cite{tanaka2019recent}, including photonics \cite{van2017advances}, electrical circuits \cite{soriano2014delay}, hydrodynamic systems \cite{lu2020supervised}, and the brain \cite{suarez_learning_2021}. However, expressivity in biological networks is not static, even in the presence of fixed weights. Instead it can be modulated dynamically over short time-scales and–––similar to regions' temporal receptive windows–––this too is spatially patterned.
~\\

Previously, we discussed how the S-A axis tracks functional specialization and integration across the cortical mantle; cortical brain systems located at the bottom of the S-A axis are responsible for processing sensory/motor information while systems at the top of the S-A axis are involved in processing higher-order cognition \cite{mesulam_representation_2008}, and the brain's connectivity allows for the hierarchical flow of information across these systems \cite{parkes_asymmetric_2022, pines_development_2023, baum_modular_2017}. However, the functional roles of these different brain systems are not static. Instead the brain utilizes a complex array of neuromodulatory systems to actively reconfigure the brain's dynamic repertoire \cite{shine_computational_2021}. In turn, this neuromodulation endows a relatively static network architecture (i.e., structural connectivity) with an increased capacity for functional flexibility. Reviewing all of the brain's neuromodulatory mechanisms, and their effects on neural dynamics, is beyond the scope of this piece (see \cite{marder_cellular_2002, bucher_beyond_2011, mccormick_neuromodulation_2020, shine_computational_2021} for reviews), as these mechanisms comprise myriad cortico-cortical, cortico-subcortical, and subcortical-subcortical interactions. Here, we focus on a specific example that we believe is well positioned to be integrated into RNNs: the balance and modulation of cortical excitation and inhibition.
~\\

One fundamental neuromodulatory effect is that of dynamic changes to cortical excitation and inhibition. Neuron's in the cortex receive a complex set of excitatory and inhibitory inputs, and the ratio between these inputs (E/I ratio) plays a critical role in coordinating an action potential. Following the S-A axis \cite{sydnor_neurodevelopment_2021}, the E/I ratio varies systematically across the cortex \cite{kim_brain-wide_2017,burt_hierarchy_2018, anderson_transcriptional_2020}, leading to baseline differences in regions' dynamics and computation \cite{deco_how_2014, deco_dynamical_2021,gao_inferring_2017, sydnor_intrinsic_2023}. Moreover, incorporating regional variations to the E/I ratio into biophysical models has been shown to improve their fit to empirical functional data \cite{deco_dynamical_2021, zhang_ei_2023}, demonstrating that the E/I ratio shapes large-scale brain dynamics. However, unlike features of brain stucture that track the S-A axis \cite{sydnor_neurodevelopment_2021}, regions' baseline E/I ratio can be dynamically shifted via up- or downregulating the excitatory and inhibitory neurotransmitters of postsynaptic cells \cite{shine_computational_2021}. This regulation is achieved via multiple neurochemical pathways which can be driven exogenously---for example, via pharmaceutical agents \cite{larsen_developmental_2022} or chemogenetics \cite{rocchi_increased_2022, markicevic_cortical_2020}---or endogenously, for example via the ascending noradrenergic arousal system (AAS) \cite{shine_computational_2021}. 
~\\

\begin{figure}[!ht]
\begin{center}
\includegraphics[width=\linewidth]{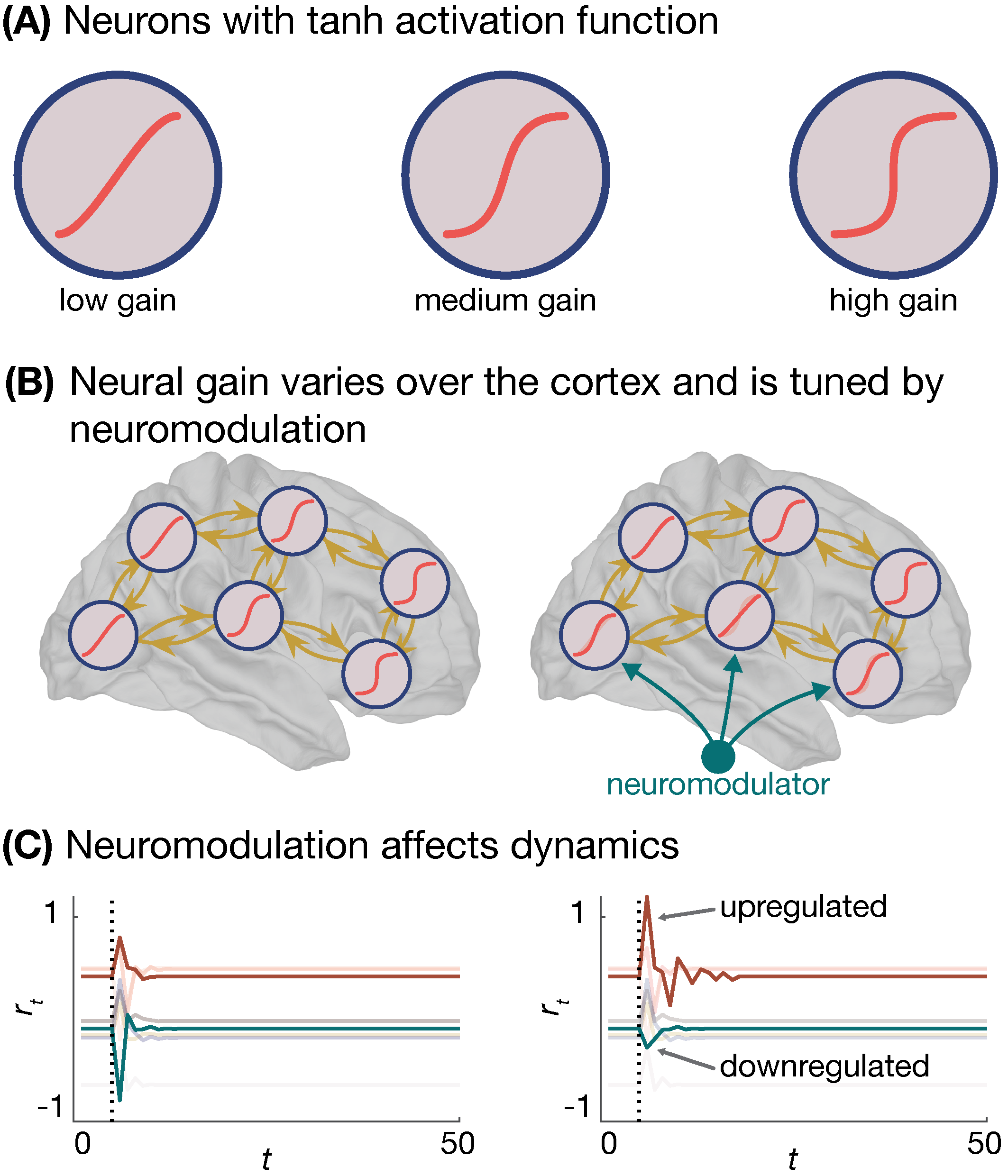}
\end{center}
\caption{\textbf{Tuning neural gain in RNNs.} \textbf{A}, Neural gain can be dynamically tuned in a trained RNN by modifying the slope of neurons' activation function. \textbf{B}, Neuroscientific research demonstrates that neural gain varies systematically across the cortex (left) and is tuned dynamically by neuromodulators (right). \textbf{C}, This neuromodulation can be used to change the dynamics of RNNs.}
\label{fig:figure3_neuromodulation}
\end{figure}

In dynamical systems, changes to neuronal excitation and inhibition are thought to engender population-level changes in neural gain (Fig. \ref{fig:figure3_neuromodulation}) \cite{shine_computational_2021}; the slope of a function that maps simulated neurons' inputs to their outputs. By tuning the neural gain between coupled oscillators, Shine \emph{et al.} \cite{shine_modulation_2018} observed that increased gain lead to greater functional integration between neural populations. Critically, functional integration is thought to be an important computational property of the brain; in the human brain, functional integration fluctuates over short time scales \cite{shine_human_2019} and facilitates cross-talk between the brain’s many functionally-specialized communities \cite{bertolero_modular_2015}. Thus, on-the-fly changes to neurons' E/I ratio facilitates a diverse range of dynamic behaviors \cite{wilson1972excitatory}. This diversity allows brain function to flexibly decouple from its underlying structural architecture \cite{vazquez-rodriguez_gradients_2019, preti_decoupling_2019, misic_network-level_2016, zamani_esfahlani_local_2022}, which in turn supports a broader range of computations than would otherwise be possible.
~\\

The above data suggests that modulation of regions' E/I ratio gives rise to state-dependent dynamics that facilitate the brain's computational expressivity. Recently, researchers have begun examining how E/I modulation might be instantiated in RNNs, with a particular focus on the aforementioned AAS. The AAS stems from the locus coeruleus, a small brainstem structure that provides diffuse noradrenergic projections spanning the cerebral cortex \cite{samuels_functional_2008, shine_computational_2021}. These projections modulate neuronal excitability via the neurotransmitter noradrenaline, granting the AAS the capacity to modulate the E/I ratio. Drawing on this mechanism, Wainstein \emph{et al.} \cite{wainstein_gain_2023} trained an RNN to perform a perceptual switching task, wherein one visual stimuli (a plane) gradually morphed into another (a shark) and the RNN was tasked with reporting which stimuli it perceived at each time point. Once trained, Wainstein \emph{et al.} \cite{wainstein_gain_2023} modified the slope of the artificial neurons' activation function (i.e., the neural gain) and examined the corresponding change in perceptual switching. The authors observed that higher gain caused perceptual switches to occur earlier than expected, while lower gain caused the opposite. Additionally, Wainstein \emph{et al.} \cite{wainstein_gain_2023} modeled the energy landscape of the RNN state-space and observed that increasing neurons' gain flattened the landscape, allowing for easier state transitions (perceptual switches). Finally, the authors supported these modeling results with task-based fMRI data as well as pupillometry data, which is thought to be an indirect measure of noradrenaline-mediated arousal \cite{joshi_pupil_2020}. Together, the authors' results demonstrate that a system's computational function can be dynamically modulated in behaviorally meaningful ways, and that this reconfiguration may be underpinned by an internal capacity to regulate neural excitability. Critically, this dynamic reconfiguration unfolds on top of a static network architecture, wherein only neurons' activation functions are tweaked while their trained weights are preserved.
~\\

The results of Wainstein \emph{et al.} \cite{wainstein_gain_2023} demonstrate that the affects of neuromodulation can be introduced to RNNs, modifying their functional outcomes in behaviorally meaningful ways. However, as touched on above, the brain comprises multiple neuromodulatory systems that are capable of influencing regions' excitation and inhibition, each of which subserve different functional goals \cite{shine_computational_2021} and each exhibit unique spatial patterning of their associated neurotransmitters and genes \cite{burt_hierarchy_2018}. Future work examining how each of these neurotransmitter maps affect RNN behavior, across a diverse range of tasks, will be important to characterize how different neuromodulatory mechanisms influence expressivity. Indeed, other fields of dynamical systems (e.g. linear systems) have already begun pursuing these goals \cite{singleton_receptor-informed_2022, luppi_transitions_2023,aitken2023neural}.

\section{Computing with the latent spaces of RNNs via constrained connectivity} \label{latent}

While expressivity tells us what information about the input is contained in a specific state, it does not tell us about the computational meaning behind that state. Specifically, although Eq.~\ref{eq:recursion} provides us with a map of how any input series $\bm{u}_\tau$ is expressed as a specific neural state $\bm{x}_{t+1}$, the meaning of that state depends on the context of the problem being solved. For example, while the dynamics of the transistors in a microprocessor can be known and simulated, the computational meaning of the transistor state depends on its internal, or \textit{latent} representation \cite{jonas2017could}. 
~\\

\begin{figure}[!ht]
\begin{center}
\includegraphics[width=\linewidth]{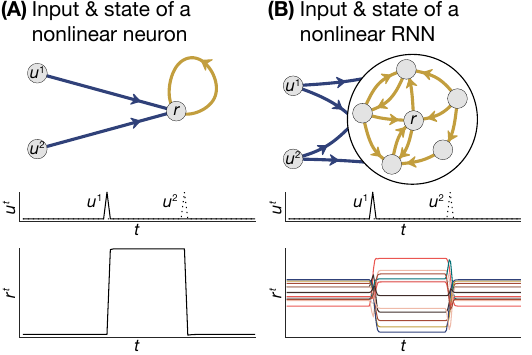}
\end{center}
\caption{\textbf{Distributed latent-space representations.} {\textbf{A}}, A nonlinear single-neuron model with two inputs (top), which has been designed to act as a memory circuit such that an impulse from input $u_1$ or $u_2$ will cause the state of the neuron to stay fixed at a high or low value, respectively (bottom). {\textbf{B}}, A nonlinear multi-neuron RNN which has been designed to act as a memory circuit (top), except that the RNN represents the states ``high'' and ``low'' as two stable fixed-points across all of its states, and these fixed-points are determined in a distributed manner \textit{via} all of the connections between the neurons. }
\label{fig:figure4_latent}
\end{figure}

As an illustration of latent representation, consider one of the fundamental memory elements of computers, the \textit{set-reset latch} (SR-latch) \cite{torii2016asic}, which simply remembers which of two inputs were pulsed most recently. A single nonlinear neuron with two inputs can be designed to mimic this behavior (Fig.~\ref{fig:figure4_latent}A), where its state remains high if input $u_1$ was last pulsed, and remains low if input $u_2$ was last pulsed. Here, the state of the neuron is directly the output of a latch. Alternatively, this latch functionality can be defined in a \textit{distributed} manner into a system of multiple neurons, where the high state is represented as some pattern of activity $\bm{r}^*$, the low state is represented as another pattern of activity $\bm{r}^\dagger$, and the input pulses transition the RNN state between these two (Fig.~\ref{fig:figure4_latent}B). Here, no single neuron is responsible for the latch dynamics. Rather, these latent-space latch dynamics depend on the connectivity between neurons, as well as how that connectivity was formed by training. In this section, we discuss how RNNs represent and manipulate information in their latent space, and the consequence of biological constraints on these latent representations. Then, we draw on recent advances from the field of neurodevelopment to put forth new directions for studying biologically-constrained RNNs.

\subsection{Sparsity and attractor stability}

Neural networks harness the power of internal, or \textit{latent}, representations for computational tasks such as path integration \cite{samsonovich1997path}, tracking \cite{fung2010moving}, and spatial working memory \cite{wimmer2014bump}. In RNNs and dynamical systems, these latent representations are often referred to as \textit{attractors}: sets of points, $\mathcal{S} = \{\bm{s}_i\}$, to which the dynamics evolve after a relatively long period of time. An early example of latent representations is associative memory in Hopfield networks \cite{hopfield1982neural}, wherein a specific set of neural activity patterns, $\mathcal{S} = \{\bm{s}_i\}$, are stored as memories that could represent information such as an image. Specifically, $\mathcal{S}$ are stored as \textit{fixed-point attractors} \cite{strogatz2018nonlinear} such that after stimulating neurons close to a specific pattern, $\bm{x}_0 = \bm{s}_i + \bm{\epsilon}$, the neural states will evolve toward a stored memory, $\bm{x}_{t\rightarrow \infty} \approx \bm{s}_i$. In general, a fixed-point is a state $\bm{x}^*$ such that
\begin{align}
\label{eq:fixed_point}
\bm{x}^* = f(\bm{x}^*).
\end{align}
The computational power of this Hopfield network is in using inputs to retrieve pre-defined information (i.e., \emph{memories}) stored in attractors, and substantial work has gone into improving their computational capability \cite{ramsauer2020hopfield} and biophysical realism \cite{storkey1997increasing}. Hence, fixed-point attractors are latent dynamical properties that can be harnessed for computation.
~\\

In addition to discrete memory states, RNNs can make use of the \textit{geometry} of their attractors to form representations and make decisions. For example, \textit{continuous-attractor neural networks} (CANNs) extend the concept of an attracting point to higher-dimensional manifolds, thereby forming attracting curves and surfaces such that the geometric position along these manifolds holds latent computational meaning. For example, the geometric trajectory of the neural network state along these manifolds can reflect a path traversed in real physical space by an agent, \cite{samsonovich1997path}, the continuous tracking of a moving stimulus \cite{fung2010moving}, and the recall of spatial location in the prefrontal cortex \cite{wimmer2014bump}. While the connectivity and dynamics of CANNs are precisely engineered to preserve translation-invariance along their structure, this continuum of attractors also emerges in trained RNN models \cite{smith2022learning}, and even in models of the prefrontal cortex trained to integrate information given different contexts \cite{mante2013context}. Hence, rather than the activity of one or a collection of specific neurons \cite{nichols2002middle}, it is the geometry of the attractor manifold that forms the internal representation of information in the RNN, and the RNN integrates external information by moving its representation along that manifold \cite{jazayeri2021interpreting}.
~\\

Formation of attractors occurs as a consequence of a loss of energy in the system, which in turn results in the stabilization of dynamics \cite{howarth_updated_2012}. This stabilization can be characterized using methods such as Lyapunov functions---energy-like quantities that monotonically decrease or dissipate throughout the dynamics \cite{giesl2015review,bellman1962vector}---and Lyapunov exponents---the rate of convergence towards an attracting manifold \cite{wolf1985determining}---to study existing systems. However, what is less clear is how connectivity can develop to improve the stabilization of these attractors. Intuitively, this energy dissipation usually takes the form of a loss in the neural \textit{activity} (e.g. the ``leaky'' component of the leaky integrator in Eq.~\ref{eq:integrator_discrete}), whereas the parameters that can be learned in the RNN are the \textit{connectivity}. Thus, the question becomes: how do we modify the RNN connectivity to achieve greater attractor stability?
~\\

\begin{figure}[!ht]
\begin{center}
\includegraphics[width=\linewidth]{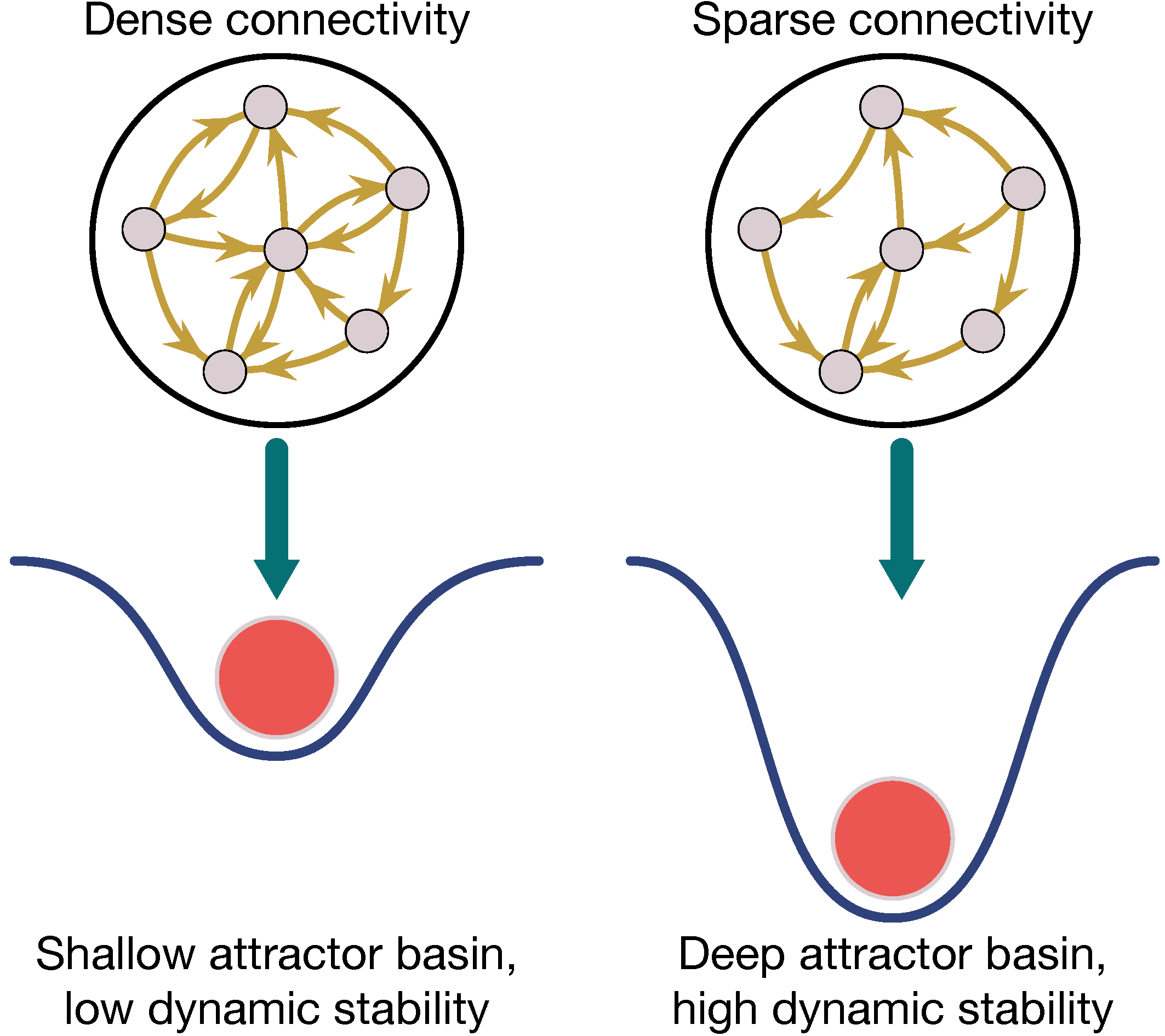}
\end{center}
\caption{\textbf{Sparse connectivity leads to stable attractors.} Pruning connectivity weights in RNNs leads to sparse weight matrices that deepen the attractor basins of RNNs dynamics, making dynamics more stable.}
\label{fig:figure5_sparisty_attractor_stability}
\end{figure}

A biologically-inspired optimization process that has proven useful for stabilizing RNN dynamics is \emph{sparseness}. In order to minimize energy expenditure \cite{howarth_updated_2012}, the brain substantially prunes its connectivity \cite{petanjek_extraordinary_2011} retaining only a sparse set of weights that are finely tuned to achieve its functional goals. In RNNs, inducing sparseness via weight pruning has been shown to provide several computational benefits. For example, Averbeck \cite{averbeck_pruning_2022} trained RNNs with and without weight pruning to complete a working memory task. Compared to their unpruned counterparts, moderate amounts of pruning yielded RNNs that (i) exhibited better task performance; (ii) required fewer training epochs; (iii) had stronger connectivity weights; and (iv) were more resistant to task distractors \cite{averbeck_pruning_2022}. Notably, regarding distractor resistance, pruned RNNs showed a smaller departure from their dynamic trajectories when they were perturbed by a distracting probe within the task. This result demonstrates that the sparse connectivity in the pruned RNNs strengthened their attractor basins (Fig. \ref{fig:figure5_sparisty_attractor_stability}), making them more stable and resistant to undesired inputs.

\subsection{Deduction: Learning Problem Structure through Iterative Algorithms}

While the geometry of dynamical attractors can be used for computational purposes, they also arise as the solutions to complex problems. For example, iterative methods are commonly used in optimization problems such as iterative refinement, \cite{moler1967iterative}, root-finding methods \cite{ypma1995historical}, and feasibility problems \cite{aragon2020douglas}. Critically, in these and many other examples, the solution is not learned in the typical deep learning sense (i.e., training), but rather emerges as an attractor to satisfy the conditions of an iterative \textit{algorithm}. In the same manner, RNNs need not only learn the attractor structure of specific input-output relations (i.e., through training), but have the potential to encode a specific algorithm in the iteration of the neural states, such that solutions to problems (i.e., tasks) emerge as attractors. 
~\\

Biological neural networks demonstrate the ability to run iterations within their latent representations \cite{flesch2022orthogonal}. A prominant example is hippocampal replay, whereby hippocampal place cells will reactivate along the same sequence as in a prior navigation experience \cite{gillespie2021hippocampal}, even when the subject is not actively performing a navigation task. Another prominant example is dynamical inference, whereby neural activity in the dlPFC can reliably predict the future nonlinear trajectory of a ball, and RNN models which best replicate prediction behavior are trained on the sequence of the ball's trajectory \cite{rajalingham2022dynamic,rajalingham2022recurrent}. Hence, RNNs are not only capable of learning attractor geometries, but also of learning and simulating the sequence of the problem structure, which may enable more generalizable solutions. 
~\\

\begin{figure}[!ht]
\begin{center}
\includegraphics[width=\linewidth]{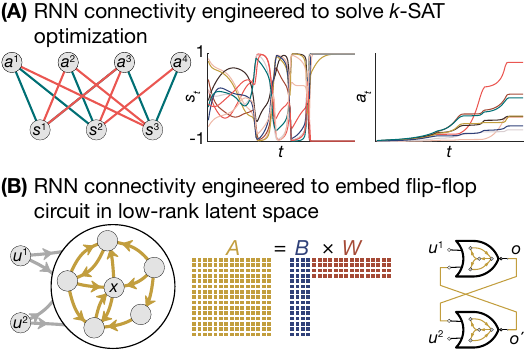}
\end{center}
\caption{\textbf{Programming Algorithms into RNNs} {\textbf{A}}, Schematic of an RNN whose neurons $s_i$ and $a_j$ represent states in a $k$-SAT problem, and whose connectivity is defined such that the stable attracting manifold is guaranteed to be a solution (left). The dynamics of the RNN evolve in a complex manner as it searches and finds the assignment of states that satisfies the $k$-SAT problem (center, right). {\textbf{B}}, Schematic of the process for designing algorithms into RNNs, where the recurrent connectivity of the RNN (gold) can be constructed from a low-rank product of matrices ($B$,$W$). Through this construction, the evolution of the RNN states concurrently evolves the state variables of the algorithm within a low-dimensional projection. This designed connectivity was used to program and simulate the SR-latch in Fig.~\ref{fig:figure4_latent}B).}
\label{fig:figure6_algorithm}
\end{figure}

RNNs can be engineered to run iterative deductions through several means. One approach involves assigning to each neuron the state variable of an algorithm, and defining complex interaction dynamics such that the RNN state will settle on the solution as a stable attractor. For example, an RNN can be designed to solve $k$-satisfiability problems, which seek an assignment of $n$ Boolean variables that satisfy a set of $c$ constraints, where each constraint places a condition on subsets of $k$ Boolean variables (Fig.~\ref{fig:figure6_algorithm}A) \cite{ercsey2011optimization}. These RNNs evolve until the neural states find a solution \cite{ercsey2011optimization}. Surprisingly, a wide variety of different dynamics and architectures can lead to different algorithms for solving the same satisfiability problem \cite{molnar2012continuous,yamashita2019bounded}, and other difficult optimization problems such as integer linear programming feasibility \cite{li2019continuous} or the $n$-queens problem \cite{ding2010high}. Hence, algorithm variables can be directly represented by individual neurons, and the algorithm rules can be directly encoded in the connectivity and update rules $f$ of RNNs. 
~\\

In addition to a direct, one-to-one encoding of problem structure as an RNN, we can also design algorithms into the latent spaces of RNNs \cite{mastrogiuseppe2018linking}. Rather than ascribing each neuron a specific variable in an algorithm, we can embed an algorithm into the distributed connectivity of an RNN (Fig.~\ref{fig:figure6_algorithm}B). One technique for this embedding is the \textit{neural engineering framework} (NEF) which enables the design of iterations of the latent-space variables through the engineering of low-rank connectivity matrices \cite{eliasmith2003neural,dewolf2020nengo}. Extensions enable the programming of iterations in pre-existing, higher-rank connectivity, and the ability to reverse-engineer representations from conventionally trained RNNs \cite{kim2023neural}. Other engineered architectures such as the differentiable neural computer \cite{graves2016hybrid} and the neural turing machine \cite{tkavcik2016neural} emulate the structure of conventional computers using differentiable neural elements. Hence, RNNs have the capability to explicitly run complex algorithms in their latent spaces, which is crucial for generalizable computation; rather than learning individual solutions, we posit that training RNNs on solution sequences will enable them to learn and generalize problem-solving strategies \cite{fawzi2022discovering}, even into nonlinear, out-of-sample regimes \cite{kim2021teaching}. 
~\\

While the above approaches may provide more generalizable RNN solutions than task-specific training, the added computational capabilities of engineered RNNs is accompanied by a further deviation of the corresponding connectivity from biology. Whether through the enforcing of low-rank connectivity \cite{eliasmith2003neural,valente2022probing} or the segregation of memory and processing units in a neural von Neumann architecture \cite{graves2016hybrid}, engineered neural connectivities lack many of the costs and constraints experienced by biological networks. A critical question then is how the brain formulates connectivities that permit sophisticated latent space computations while adhering to biological constraints. Prior work has demonstrated the capability of largely disordered RNNs to produce sequences that rely on recurrent connectivity \cite{rajan2016recurrent}, and the importance of the sequence of learning over many learning iterations---a curriculum---for RNN performance \cite{kepple2022curriculum}. Hence, rather than forming low-rank structures \textit{ab initio}, the brain defines its structure and dynamics through learning and plasticity on long time scales. We explore insights into the governing principles and computational advantages of this progression in neurodevelopment.

\subsection{Latent-space computational capability throughout neurodevelopment}

\begin{figure*}[!ht]
\begin{center}
\includegraphics[width=\textwidth]{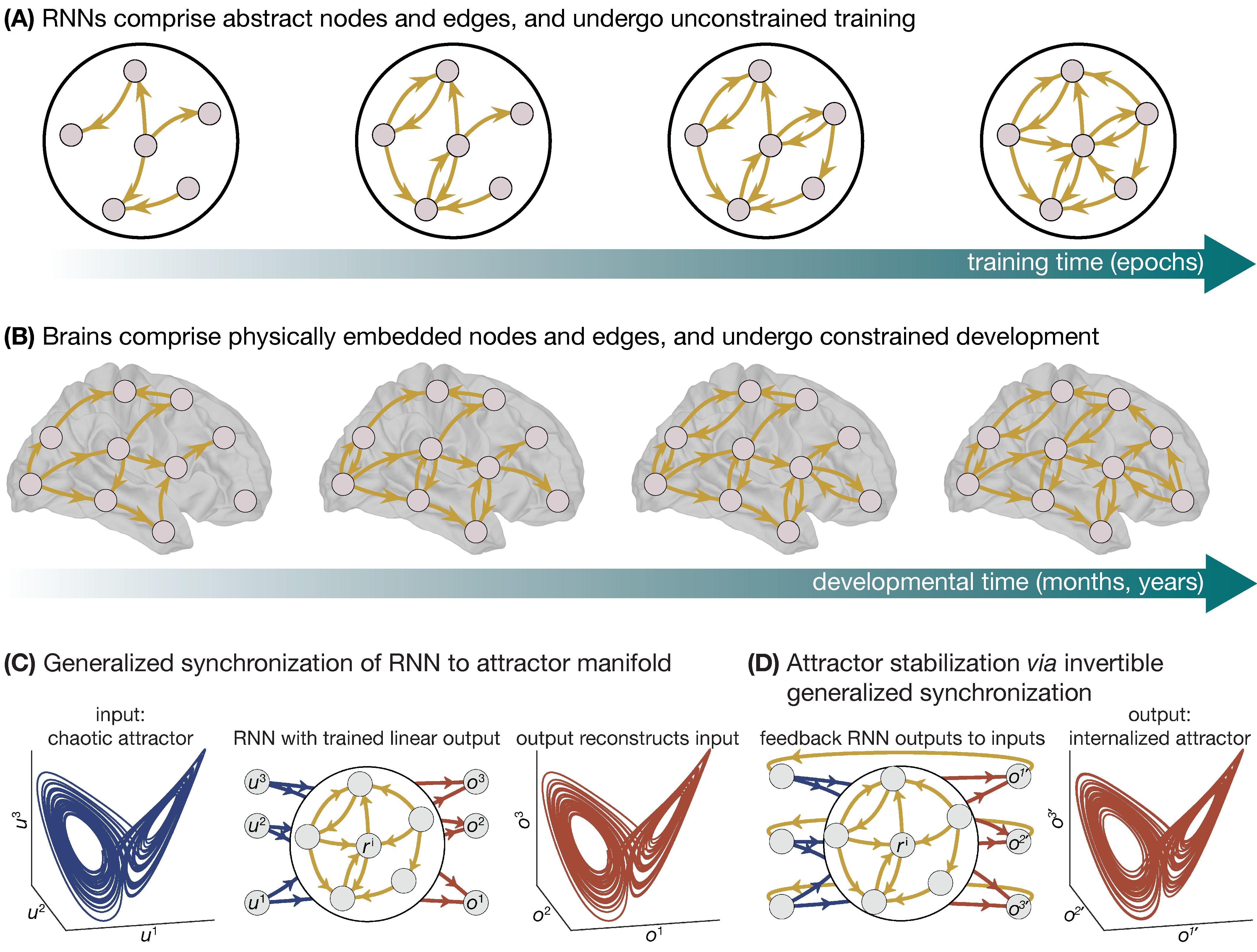}
\end{center}
\caption{\textbf{Computational capabilities of RNNs and biological brains throughout development}. RNNs and brains are subject to different environments. \textbf{A}, RNNs comprise abstract nodes and edges, and their training is unconstrained. This allows RNNs to evolve toward any network architecture that optimally performs a given task. \textbf{B}, By contrast, brains comprise physically embedded nodes and edges, and the formations of their network architecture is tightly constrained by space and time. \textbf{C}, The extent to which pre-existing connectivities can \textit{follow} latent representations can be captured through \textit{generalized synchronization}, which measures how well an RNN that is driven by an attracting manifold (in this case, the chaotic Lorenz attractor) can reconstruct the driving signal from its states. \textbf{D}, The ability of pre-existing connectivities to \textit{internalize} and generate latent representations is captured through \textit{invertible generalized synchronization}, whereby an RNN that has learned output weights $W$ that copy the attractor manifold can then \textit{drive itself} by feeding those outputs back as inputs to autonomously generate the attractor \textit{without any input}.}
\label{fig:figure7_neurodevelopment}
\end{figure*}

Just as the topology of an RNN is sculpted over training epochs (Fig. \ref{fig:figure7_neurodevelopment}A), the topology of the human connectome is sculpted throughout development (Fig. \ref{fig:figure7_neurodevelopment}B). However, unlike the RNN, which may be trained in an unconstrained manner, neurodevelopment follows a carefully orchestrated and stereotyped program that unfolds dynamically across space and time. Specifically, cortical neurodevelopment is thought to spatially track the aforementioned S-A axis in a temporally staged manner \cite{sydnor_neurodevelopment_2021, larsen_critical_2023}, and this staging is thought to underpin the emergence of cortical regions' functional specialization and inter-connectivity. Crucially, the asynchronous nature of this developmental program is thought to underpin the sequential emergence of increasingly complex cognitive functions \cite{kim_biased_2019, tervo-clemmens_canonical_2022}, suggesting that neurodevelopment stages the brain's acquisition of lower- and higher-order computational processes. Mechanistically, this program may be underpinned by windows of heightened neural plasticity that cascade up the S-A axis \cite{larsen_adolescence_2018, larsen_critical_2023} priming specific neural circuits at specific points in time for experience-dependent neural change.
~\\

Regions in the cortex are defined in part according their laminar structure \cite{garcia-cabezas_structural_2019,garcia-cabezas_protocol_2020}, with different regions exhibiting variations in the number and size of their distinct layers, as well as different distributions of cells throughout those layers. Critically, cortical variations in cytoarchitecture conform to the S-A axis \cite{paquola_microstructural_2019}, and animal research demonstrates that this spatial patterning predicts regions' extrinsic connectivity \cite{barbas_general_2015}, including their strength \cite{markov_anatomy_2014}, distance \cite{markov_anatomy_2014}, and layer-wise projections \cite{markov_anatomy_2014, beul_towards_2015}. In humans, structural connectivity between regions at the bottom of the S-A axis refines relatively early in development, while connectivity at the top of the S-A axis does so later in development \cite{huttenlocher_synaptogenesis_1982, peter_r_synaptic_1979, semple_brain_2013, buckner_evolution_2013, baum_modular_2017}. Furthermore, recent work has shown that cytoarchitecture plays an important role in shaping how dynamics spread across the connectome throughout development \cite{parkes_asymmetric_2022}. Thus, the spatial patterning embedded in the S-A axis plays a key role in shaping connectome topology throughout development. But what about RNNs? Recent work by Achterberg \emph{et al.} \cite{achterberg_spatially_2023} regularized RNNs by using the Euclidean distance between regions to constrain training. The authors found that modularity \cite{sporns_modular_2016} and small-worldness \cite{bassett_small-world_2017}---two complex topological features that are hallmarks of the human connectome \cite{bullmore_economy_2012}---emerged to a greater extent in these spatially-embedded RNNs compared to standard RNNs (see also recent work by Tanner \emph{et al.} \cite{tanner_functional_2023} for evidence of modularity in RNNs trained without spatial constraints). Additionally, this effect coincided with achieving higher out-of-sample task performance earlier in training compared to standard RNNs (though performance eventually converged; see their Figure 2A).
~\\

The results of Achterberg \emph{et al.} \cite{achterberg_spatially_2023} demonstrate that incorporating space-based inductive biases into RNN training causes them to converge on topological features observed in the human connectome. However, it remains unclear whether neurodevelopmentally-informed spatial constraints, like those embedded in the S-A axis \cite{sydnor_neurodevelopment_2021}, show similar effects. We posit that constraining RNNs using the S-A axis may outperform Euclidean distance-based spatial embedding, as the former is rooted in evolutionary programs of connectivity formation and functional specialization \cite{xu_cross-species_2020}. Additionally, the spatial constraints deployed by Achterberg \emph{et al.} \cite{achterberg_spatially_2023} were static throughout training. As mentioned above, the S-A axis scaffolds connectome development in a temporally varying way \cite{larsen_critical_2023}, and incorporating this dynamic information will be critical to achieving realistic brain-like topology in RNNs. One approach would be to code spatially varying periods of heightened learning into RNNs, simulating traveling waves of heightened neural plasticity \cite{larsen_adolescence_2018, larsen_critical_2023}. Such an inductive bias could be achieved by including temporal cascades of weight training that flow bottom-up across the S-A axis.
~\\

In addition to injecting spatially constrained inductive biases into RNNs, we can also directly assess the ability of the developing connectome to support latent-space computation. This analysis can be achieved through studying the \textit{synchronization} between a given RNN and a particular latent attractor. Historically, synchronization has been shown to be crucial for computation in both biological cortical networks \cite{fries2009neuronal} and artifical RNNs \cite{loebel2002computation}, and is deeply related to consensus dynamics \cite{li2009consensus}. Intuitively, synchronization between two systems implies that both systems are evolving identically. This concept can be extended to \textit{generalized synchronization}, which stipulates conditions under which a response system, $\bm{r}_t$, has synchronized in a general sense to a driving system, $\bm{d}_t$ \cite{rulkov1995generalized}, such that rather than evolving identically such that $\bm{r}_t = \bm{d}_t$ \cite{pecora1990synchronization}, the joint system has collapsed onto a function $\phi$ of only the driver system such that $\bm{r}_t = \phi(\bm{d}_t)$. Under these conditions, the response system has \textit{followed} the attractor structure of the driving system. If we choose the response system to be an RNN with an empirically-derived connectivity taken at a specific point in neurodevelopment, and the driver system to be a specific latent attractor dynamical system, then we can assess whether the RNN can follow the attractor structure of the driving system (Fig.~\ref{fig:figure7_neurodevelopment}C). 
~\\

Of course, just because the RNN can synchronize to a particular latent-space attractor does not guarantee that it can maintain that attractor once the driving system is gone. In order for the RNN to \textit{internalize} the attractor dynamics, theories of \textit{invertible generalized synchronization} (IGS) stipulate conditions for which the attractor of the driver system can be invertibly reproduced and stabilized by the response system, and thus can be \textit{learned} autonomously \cite{lu2020invertible}. Hence, rather than modifying RNN connectivity to learn a latent space conditioned on spatial and resource constraints, IGS tests whether a given RNN connectivity that already obeys those spatial and resource constraints can stably generate a latent attractor (Fig.~\ref{fig:figure7_neurodevelopment}D). The IGS theory also indicates that the ability to internalize latent attractors from driving systems depends not only on the RNN's structure, but also the latent attractor. Thus, given RNNs of different structures throughout development, we may examine the IGS on their most-likely encountered driving signals corresponding to their position along the S-A axis. 

\section{Conclusions}
Neural systems compute using dynamics, and the dynamics of biological brains evolve atop the computational substrate of a spatially and resource constrained network. Here, we sought to jointly discuss advances in neuroscience and dynamical systems with a view to improving the computational power and interpretability of RNNs. Within the context of two computational capabilities---expressivity and latent-space computing---we highlighted several avenues for future research that we believe will advance our understanding of dynamical computation. Through these avenues, we envision biologically interpretable, computationally improved RNN models of how the brain computes.
~\\

Central to this research program is the application of biophysical constraints on the computational dynamics of RNNs. While not exhaustive, the constraints discussed herein represent the diverse influences of neurobiology on network structure, and they have been shown to influence the emergence of complex behavior in humans. Furthermore, the influence of these constraints on RNN connectivity can studied in combination. For example, the spatial-patterning of regions' baseline E/I ratio emerges throughout development \cite{larsen_critical_2023}, indicating it's connection to the S-A axis. Another example is to explore how to extend low-rank RNN design approaches to higher-rank connectivities that more closely match the spatial gradients of the S-A axis. Thus, these biophysical constraints provide fertile ground for future experimental, computational, and theoretical work into biologically-informed RNNs.
~\\

The methodological strategies for incorporating these constraints into computational RNN modeling are vast. Here, we have discussed several approaches: using known cortical structure and function as target connectivities of additional training constraints \cite{sydnor_neurodevelopment_2021,achterberg_spatially_2023}, using resource constraints as an alternative learning mechanism \cite{ali2022predictive}, dynamically altering connectivity with neuromodulation for increased expressivity \cite{shine_modulation_2018}, stabilizing latent attractors through pruning \cite{averbeck_pruning_2022}, engineering and modifying low-rank latent representations and algorithms \cite{eliasmith2003neural,kim2023neural}, probing the teachability of RNNs \textit{via} synchronization \cite{lu2020invertible}, among many others. Hence, there is no one prescription that serves as a panacea for the rich problems that lie at the intersection of computation, dynamics, and neurobiology. Instead, we must continue developing diverse and creative approaches for maximizing the computational capabilities of neurodynamical models through the use of biological and developmental constraints. 

\section*{Acknowledgements}
We gratefully acknowledge comments and conversations with Dr. Zhixin Lu, Dr. Harang Ju, Dr. Dale Zhou, Dr. Melody Lim, and Dr. Stephen Hanson. JZK was supported by the Bethe/KIC/Wilkins postdoctoral fellowship. BL was supported by the National Institute Of Mental Health of the National Institutes of Health under Award Number R00MH127293. LP was supported by the National Institute Of Mental Health of the National Institutes of Health under Award Number R00MH127296.


\section*{References}
\begin{thebibliography}{100}
	\expandafter\ifx\csname url\endcsname\relax
	\def\url#1{\texttt{#1}}\fi
	\expandafter\ifx\csname urlprefix\endcsname\relax\def\urlprefix{URL }\fi
	\providecommand{\bibinfo}[2]{#2}
	\providecommand{\eprint}[2][]{\url{#2}}
	
	\bibitem{ijspeert2008central}
	\bibinfo{author}{Ijspeert, A.~J.}
	\newblock \bibinfo{title}{Central pattern generators for locomotion control in
		animals and robots: a review}.
	\newblock \emph{\bibinfo{journal}{Neural networks}}
	\textbf{\bibinfo{volume}{21}}, \bibinfo{pages}{642--653}
	(\bibinfo{year}{2008}).
	
	\bibitem{seeley2007dissociable}
	\bibinfo{author}{Seeley, W.~W.} \emph{et~al.}
	\newblock \bibinfo{title}{Dissociable intrinsic connectivity networks for
		salience processing and executive control}.
	\newblock \emph{\bibinfo{journal}{Journal of Neuroscience}}
	\textbf{\bibinfo{volume}{27}}, \bibinfo{pages}{2349--2356}
	(\bibinfo{year}{2007}).
	
	\bibitem{mante2013context}
	\bibinfo{author}{Mante, V.}, \bibinfo{author}{Sussillo, D.},
	\bibinfo{author}{Shenoy, K.~V.} \& \bibinfo{author}{Newsome, W.~T.}
	\newblock \bibinfo{title}{Context-dependent computation by recurrent dynamics
		in prefrontal cortex}.
	\newblock \emph{\bibinfo{journal}{nature}} \textbf{\bibinfo{volume}{503}},
	\bibinfo{pages}{78--84} (\bibinfo{year}{2013}).
	
	\bibitem{werbos1990backpropagation}
	\bibinfo{author}{Werbos, P.~J.}
	\newblock \bibinfo{title}{Backpropagation through time: what it does and how to
		do it}.
	\newblock \emph{\bibinfo{journal}{Proceedings of the IEEE}}
	\textbf{\bibinfo{volume}{78}}, \bibinfo{pages}{1550--1560}
	(\bibinfo{year}{1990}).
	
	\bibitem{betzel_generative_2016}
	\bibinfo{author}{Betzel, R.~F.} \emph{et~al.}
	\newblock \bibinfo{title}{Generative models of the human connectome}.
	\newblock \emph{\bibinfo{journal}{NeuroImage}} \textbf{\bibinfo{volume}{124}},
	\bibinfo{pages}{1054--1064} (\bibinfo{year}{2016}).
	
	\bibitem{oldham_development_2019}
	\bibinfo{author}{Oldham, S.} \& \bibinfo{author}{Fornito, A.}
	\newblock \bibinfo{title}{The development of brain network hubs}.
	\newblock \emph{\bibinfo{journal}{Developmental Cognitive Neuroscience}}
	\textbf{\bibinfo{volume}{36}}, \bibinfo{pages}{100607}
	(\bibinfo{year}{2019}).
	
	\bibitem{oldham_modeling_2022}
	\bibinfo{author}{Oldham, S.} \emph{et~al.}
	\newblock \bibinfo{title}{Modeling spatial, developmental, physiological, and
		topological constraints on human brain connectivity}.
	\newblock \emph{\bibinfo{journal}{Science Advances}}
	\textbf{\bibinfo{volume}{8}}, \bibinfo{pages}{eabm6127}
	(\bibinfo{year}{2022}).
	
	\bibitem{akarca_generative_2021}
	\bibinfo{author}{Akarca, D.} \emph{et~al.}
	\newblock \bibinfo{title}{A generative network model of neurodevelopmental
		diversity in structural brain organization}.
	\newblock \emph{\bibinfo{journal}{Nature Communications}}
	\textbf{\bibinfo{volume}{12}}, \bibinfo{pages}{4216} (\bibinfo{year}{2021}).
	
	\bibitem{hodgkin1952quantitative}
	\bibinfo{author}{Hodgkin, A.~L.} \& \bibinfo{author}{Huxley, A.~F.}
	\newblock \bibinfo{title}{A quantitative description of membrane current and
		its application to conduction and excitation in nerve}.
	\newblock \emph{\bibinfo{journal}{The Journal of physiology}}
	\textbf{\bibinfo{volume}{117}}, \bibinfo{pages}{500} (\bibinfo{year}{1952}).
	
	\bibitem{petousakis2022impact}
	\bibinfo{author}{Petousakis, K.-E.}, \bibinfo{author}{Apostolopoulou, A.~A.} \&
	\bibinfo{author}{Poirazi, P.}
	\newblock \bibinfo{title}{The impact of hodgkin--huxley models on dendritic
		research}.
	\newblock \emph{\bibinfo{journal}{The Journal of Physiology}}
	(\bibinfo{year}{2022}).
	
	\bibitem{wilson1972excitatory}
	\bibinfo{author}{Wilson, H.~R.} \& \bibinfo{author}{Cowan, J.~D.}
	\newblock \bibinfo{title}{Excitatory and inhibitory interactions in localized
		populations of model neurons}.
	\newblock \emph{\bibinfo{journal}{Biophysical journal}}
	\textbf{\bibinfo{volume}{12}}, \bibinfo{pages}{1--24} (\bibinfo{year}{1972}).
	
	\bibitem{pinto1996quantitative}
	\bibinfo{author}{Pinto, D.~J.}, \bibinfo{author}{Brumberg, J.~C.},
	\bibinfo{author}{Simons, D.~J.}, \bibinfo{author}{Ermentrout, G.~B.} \&
	\bibinfo{author}{Traub, R.}
	\newblock \bibinfo{title}{A quantitative population model of whisker barrels:
		re-examining the wilson-cowan equations}.
	\newblock \emph{\bibinfo{journal}{Journal of computational neuroscience}}
	\textbf{\bibinfo{volume}{3}}, \bibinfo{pages}{247--264}
	(\bibinfo{year}{1996}).
	
	\bibitem{sadeghi2020dynamic}
	\bibinfo{author}{Sadeghi, S.}, \bibinfo{author}{Mier, D.},
	\bibinfo{author}{Gerchen, M.~F.}, \bibinfo{author}{Schmidt, S.~N.} \&
	\bibinfo{author}{Hass, J.}
	\newblock \bibinfo{title}{Dynamic causal modeling for fmri with
		wilson-cowan-based neuronal equations}.
	\newblock \emph{\bibinfo{journal}{Frontiers in Neuroscience}}
	\textbf{\bibinfo{volume}{14}}, \bibinfo{pages}{593867}
	(\bibinfo{year}{2020}).
	
	\bibitem{parkes_using_2023}
	\bibinfo{author}{Parkes, L.} \emph{et~al.}
	\newblock \bibinfo{title}{Using network control theory to study the dynamics of
		the structural connectome}.
	\newblock \bibinfo{type}{preprint}, \bibinfo{institution}{Neuroscience}
	(\bibinfo{year}{2023}).
	
	\bibitem{lynn_physics_2019}
	\bibinfo{author}{Lynn, C.~W.} \& \bibinfo{author}{Bassett, D.~S.}
	\newblock \bibinfo{title}{The physics of brain network structure, function and
		control}.
	\newblock \emph{\bibinfo{journal}{Nature Reviews Physics}}
	\textbf{\bibinfo{volume}{1}}, \bibinfo{pages}{318--332}
	(\bibinfo{year}{2019}).
	
	\bibitem{seguin_brain_2023}
	\bibinfo{author}{Seguin, C.}, \bibinfo{author}{Sporns, O.} \&
	\bibinfo{author}{Zalesky, A.}
	\newblock \bibinfo{title}{Brain network communication: concepts, models and
		applications}.
	\newblock \emph{\bibinfo{journal}{Nature Reviews Neuroscience}}
	(\bibinfo{year}{2023}).
	
	\bibitem{srivastava_models_2020}
	\bibinfo{author}{Srivastava, P.} \emph{et~al.}
	\newblock \bibinfo{title}{Models of communication and control for brain
		networks: distinctions, convergence, and future outlook}.
	\newblock \emph{\bibinfo{journal}{Network Neuroscience}}
	\textbf{\bibinfo{volume}{4}}, \bibinfo{pages}{1122--1159}
	(\bibinfo{year}{2020}).
	
	\bibitem{breakspear_dynamic_2017}
	\bibinfo{author}{Breakspear, M.}
	\newblock \bibinfo{title}{Dynamic models of large-scale brain activity}.
	\newblock \emph{\bibinfo{journal}{Nature Neuroscience}}
	\textbf{\bibinfo{volume}{20}}, \bibinfo{pages}{340--352}
	(\bibinfo{year}{2017}).
	
	\bibitem{roberts_metastable_2019}
	\bibinfo{author}{Roberts, J.~A.} \emph{et~al.}
	\newblock \bibinfo{title}{Metastable brain waves}.
	\newblock \emph{\bibinfo{journal}{Nature Communications}}
	\textbf{\bibinfo{volume}{10}}, \bibinfo{pages}{1056} (\bibinfo{year}{2019}).
	
	\bibitem{fitzhugh1961impulses}
	\bibinfo{author}{FitzHugh, R.}
	\newblock \bibinfo{title}{Impulses and physiological states in theoretical
		models of nerve membrane}.
	\newblock \emph{\bibinfo{journal}{Biophysical journal}}
	\textbf{\bibinfo{volume}{1}}, \bibinfo{pages}{445--466}
	(\bibinfo{year}{1961}).
	
	\bibitem{mcculloch1943logical}
	\bibinfo{author}{McCulloch, W.~S.} \& \bibinfo{author}{Pitts, W.}
	\newblock \bibinfo{title}{A logical calculus of the ideas immanent in nervous
		activity}.
	\newblock \emph{\bibinfo{journal}{The bulletin of mathematical biophysics}}
	\textbf{\bibinfo{volume}{5}}, \bibinfo{pages}{115--133}
	(\bibinfo{year}{1943}).
	
	\bibitem{chung2021turing}
	\bibinfo{author}{Chung, S.} \& \bibinfo{author}{Siegelmann, H.}
	\newblock \bibinfo{title}{Turing completeness of bounded-precision recurrent
		neural networks}.
	\newblock \emph{\bibinfo{journal}{Advances in Neural Information Processing
			Systems}} \textbf{\bibinfo{volume}{34}}, \bibinfo{pages}{28431--28441}
	(\bibinfo{year}{2021}).
	
	\bibitem{liu2020dstp}
	\bibinfo{author}{Liu, Y.}, \bibinfo{author}{Gong, C.}, \bibinfo{author}{Yang,
		L.} \& \bibinfo{author}{Chen, Y.}
	\newblock \bibinfo{title}{Dstp-rnn: A dual-stage two-phase attention-based
		recurrent neural network for long-term and multivariate time series
		prediction}.
	\newblock \emph{\bibinfo{journal}{Expert Systems with Applications}}
	\textbf{\bibinfo{volume}{143}}, \bibinfo{pages}{113082}
	(\bibinfo{year}{2020}).
	
	\bibitem{lu2020supervised}
	\bibinfo{author}{Lu, Z.}, \bibinfo{author}{Kim, J.~Z.} \&
	\bibinfo{author}{Bassett, D.~S.}
	\newblock \bibinfo{title}{Supervised chaotic source separation by a tank of
		water}.
	\newblock \emph{\bibinfo{journal}{Chaos: An Interdisciplinary Journal of
			Nonlinear Science}} \textbf{\bibinfo{volume}{30}} (\bibinfo{year}{2020}).
	
	\bibitem{wang_evolving_2021}
	\bibinfo{author}{Wang, P.~Y.}, \bibinfo{author}{Sun, Y.},
	\bibinfo{author}{Axel, R.}, \bibinfo{author}{Abbott, L.} \&
	\bibinfo{author}{Yang, G.~R.}
	\newblock \bibinfo{title}{Evolving the olfactory system with machine learning}.
	\newblock \emph{\bibinfo{journal}{Neuron}} \textbf{\bibinfo{volume}{109}},
	\bibinfo{pages}{3879--3892.e5} (\bibinfo{year}{2021}).
	
	\bibitem{yang_task_2019}
	\bibinfo{author}{Yang, G.~R.}, \bibinfo{author}{Joglekar, M.~R.},
	\bibinfo{author}{Song, H.~F.}, \bibinfo{author}{Newsome, W.~T.} \&
	\bibinfo{author}{Wang, X.-J.}
	\newblock \bibinfo{title}{Task representations in neural networks trained to
		perform many cognitive tasks}.
	\newblock \emph{\bibinfo{journal}{Nature Neuroscience}}
	\textbf{\bibinfo{volume}{22}}, \bibinfo{pages}{297--306}
	(\bibinfo{year}{2019}).
	
	\bibitem{roberts_contribution_2016}
	\bibinfo{author}{Roberts, J.~A.} \emph{et~al.}
	\newblock \bibinfo{title}{The contribution of geometry to the human
		connectome}.
	\newblock \emph{\bibinfo{journal}{NeuroImage}} \textbf{\bibinfo{volume}{124}},
	\bibinfo{pages}{379--393} (\bibinfo{year}{2016}).
	
	\bibitem{bullmore_economy_2012}
	\bibinfo{author}{Bullmore, E.} \& \bibinfo{author}{Sporns, O.}
	\newblock \bibinfo{title}{The economy of brain network organization}.
	\newblock \emph{\bibinfo{journal}{Nature Reviews Neuroscience}}
	\textbf{\bibinfo{volume}{13}}, \bibinfo{pages}{336--349}
	(\bibinfo{year}{2012}).
	
	\bibitem{ramon_y_cajal_histology_1995}
	\bibinfo{author}{Ramón~y Cajal, S.}
	\newblock \emph{\bibinfo{title}{Histology of the nervous system of man and
			vertebrates}}.
	\newblock No. \bibinfo{number}{no. 6} in \bibinfo{series}{History of
		neuroscience} (\bibinfo{publisher}{Oxford University Press},
	\bibinfo{address}{New York}, \bibinfo{year}{1995}).
	
	\bibitem{dupont2019augmented}
	\bibinfo{author}{Dupont, E.}, \bibinfo{author}{Doucet, A.} \&
	\bibinfo{author}{Teh, Y.~W.}
	\newblock \bibinfo{title}{Augmented neural odes}.
	\newblock \emph{\bibinfo{journal}{Advances in neural information processing
			systems}} \textbf{\bibinfo{volume}{32}} (\bibinfo{year}{2019}).
	
	\bibitem{lukovsevivcius2009reservoir}
	\bibinfo{author}{Luko{\v{s}}evi{\v{c}}ius, M.} \& \bibinfo{author}{Jaeger, H.}
	\newblock \bibinfo{title}{Reservoir computing approaches to recurrent neural
		network training}.
	\newblock \emph{\bibinfo{journal}{Computer science review}}
	\textbf{\bibinfo{volume}{3}}, \bibinfo{pages}{127--149}
	(\bibinfo{year}{2009}).
	
	\bibitem{ali2022predictive}
	\bibinfo{author}{Ali, A.}, \bibinfo{author}{Ahmad, N.},
	\bibinfo{author}{de~Groot, E.}, \bibinfo{author}{van Gerven, M. A.~J.} \&
	\bibinfo{author}{Kietzmann, T.~C.}
	\newblock \bibinfo{title}{Predictive coding is a consequence of energy
		efficiency in recurrent neural networks}.
	\newblock \emph{\bibinfo{journal}{Patterns}} \textbf{\bibinfo{volume}{3}}
	(\bibinfo{year}{2022}).
	
	\bibitem{felleisen1991expressive}
	\bibinfo{author}{Felleisen, M.}
	\newblock \bibinfo{title}{On the expressive power of programming languages}.
	\newblock \emph{\bibinfo{journal}{Science of computer programming}}
	\textbf{\bibinfo{volume}{17}}, \bibinfo{pages}{35--75}
	(\bibinfo{year}{1991}).
	
	\bibitem{sipser1996introduction}
	\bibinfo{author}{Sipser, M.}
	\newblock \bibinfo{title}{Introduction to the theory of computation}.
	\newblock \emph{\bibinfo{journal}{ACM Sigact News}}
	\textbf{\bibinfo{volume}{27}}, \bibinfo{pages}{27--29}
	(\bibinfo{year}{1996}).
	
	\bibitem{hornik1989multilayer}
	\bibinfo{author}{Hornik, K.}, \bibinfo{author}{Stinchcombe, M.} \&
	\bibinfo{author}{White, H.}
	\newblock \bibinfo{title}{Multilayer feedforward networks are universal
		approximators}.
	\newblock \emph{\bibinfo{journal}{Neural networks}}
	\textbf{\bibinfo{volume}{2}}, \bibinfo{pages}{359--366}
	(\bibinfo{year}{1989}).
	
	\bibitem{cybenko1989approximation}
	\bibinfo{author}{Cybenko, G.}
	\newblock \bibinfo{title}{Approximation by superpositions of a sigmoidal
		function}.
	\newblock \emph{\bibinfo{journal}{Mathematics of control, signals and systems}}
	\textbf{\bibinfo{volume}{2}}, \bibinfo{pages}{303--314}
	(\bibinfo{year}{1989}).
	
	\bibitem{schafer2006recurrent}
	\bibinfo{author}{Sch{\"a}fer, A.~M.} \& \bibinfo{author}{Zimmermann, H.~G.}
	\newblock \bibinfo{title}{Recurrent neural networks are universal
		approximators}.
	\newblock In \emph{\bibinfo{booktitle}{Artificial Neural Networks--ICANN 2006:
			16th International Conference, Athens, Greece, September 10-14, 2006.
			Proceedings, Part I 16}}, \bibinfo{pages}{632--640}
	(\bibinfo{organization}{Springer}, \bibinfo{year}{2006}).
	
	\bibitem{poole2016exponential}
	\bibinfo{author}{Poole, B.}, \bibinfo{author}{Lahiri, S.},
	\bibinfo{author}{Raghu, M.}, \bibinfo{author}{Sohl-Dickstein, J.} \&
	\bibinfo{author}{Ganguli, S.}
	\newblock \bibinfo{title}{Exponential expressivity in deep neural networks
		through transient chaos}.
	\newblock \emph{\bibinfo{journal}{Advances in neural information processing
			systems}} \textbf{\bibinfo{volume}{29}} (\bibinfo{year}{2016}).
	
	\bibitem{raghu2017expressive}
	\bibinfo{author}{Raghu, M.}, \bibinfo{author}{Poole, B.},
	\bibinfo{author}{Kleinberg, J.}, \bibinfo{author}{Ganguli, S.} \&
	\bibinfo{author}{Sohl-Dickstein, J.}
	\newblock \bibinfo{title}{On the expressive power of deep neural networks}.
	\newblock In \emph{\bibinfo{booktitle}{international conference on machine
			learning}}, \bibinfo{pages}{2847--2854} (\bibinfo{organization}{PMLR},
	\bibinfo{year}{2017}).
	
	\bibitem{bertschinger2004real}
	\bibinfo{author}{Bertschinger, N.} \& \bibinfo{author}{Natschl{\"a}ger, T.}
	\newblock \bibinfo{title}{Real-time computation at the edge of chaos in
		recurrent neural networks}.
	\newblock \emph{\bibinfo{journal}{Neural computation}}
	\textbf{\bibinfo{volume}{16}}, \bibinfo{pages}{1413--1436}
	(\bibinfo{year}{2004}).
	
	\bibitem{sompolinsky1988chaos}
	\bibinfo{author}{Sompolinsky, H.}, \bibinfo{author}{Crisanti, A.} \&
	\bibinfo{author}{Sommers, H.-J.}
	\newblock \bibinfo{title}{Chaos in random neural networks}.
	\newblock \emph{\bibinfo{journal}{Physical review letters}}
	\textbf{\bibinfo{volume}{61}}, \bibinfo{pages}{259} (\bibinfo{year}{1988}).
	
	\bibitem{kim2023neural}
	\bibinfo{author}{Kim, J.~Z.} \& \bibinfo{author}{Bassett, D.~S.}
	\newblock \bibinfo{title}{A neural machine code and programming framework for
		the reservoir computer}.
	\newblock \emph{\bibinfo{journal}{Nature Machine Intelligence}}
	\bibinfo{pages}{1--9} (\bibinfo{year}{2023}).
	
	\bibitem{rajan2006eigenvalue}
	\bibinfo{author}{Rajan, K.} \& \bibinfo{author}{Abbott, L.~F.}
	\newblock \bibinfo{title}{Eigenvalue spectra of random matrices for neural
		networks}.
	\newblock \emph{\bibinfo{journal}{Physical review letters}}
	\textbf{\bibinfo{volume}{97}}, \bibinfo{pages}{188104}
	(\bibinfo{year}{2006}).
	
	\bibitem{langton1990computation}
	\bibinfo{author}{Langton, C.~G.}
	\newblock \bibinfo{title}{Computation at the edge of chaos: Phase transitions
		and emergent computation}.
	\newblock \emph{\bibinfo{journal}{Physica D: nonlinear phenomena}}
	\textbf{\bibinfo{volume}{42}}, \bibinfo{pages}{12--37}
	(\bibinfo{year}{1990}).
	
	\bibitem{sussillo2009generating}
	\bibinfo{author}{Sussillo, D.} \& \bibinfo{author}{Abbott, L.~F.}
	\newblock \bibinfo{title}{Generating coherent patterns of activity from chaotic
		neural networks}.
	\newblock \emph{\bibinfo{journal}{Neuron}} \textbf{\bibinfo{volume}{63}},
	\bibinfo{pages}{544--557} (\bibinfo{year}{2009}).
	
	\bibitem{munoz2018colloquium}
	\bibinfo{author}{Munoz, M.~A.}
	\newblock \bibinfo{title}{Colloquium: Criticality and dynamical scaling in
		living systems}.
	\newblock \emph{\bibinfo{journal}{Reviews of Modern Physics}}
	\textbf{\bibinfo{volume}{90}}, \bibinfo{pages}{031001}
	(\bibinfo{year}{2018}).
	
	\bibitem{hochstetter2021avalanches}
	\bibinfo{author}{Hochstetter, J.} \emph{et~al.}
	\newblock \bibinfo{title}{Avalanches and edge-of-chaos learning in neuromorphic
		nanowire networks}.
	\newblock \emph{\bibinfo{journal}{Nature Communications}}
	\textbf{\bibinfo{volume}{12}}, \bibinfo{pages}{4008} (\bibinfo{year}{2021}).
	
	\bibitem{ju2020network}
	\bibinfo{author}{Ju, H.}, \bibinfo{author}{Kim, J.~Z.}, \bibinfo{author}{Beggs,
		J.~M.} \& \bibinfo{author}{Bassett, D.~S.}
	\newblock \bibinfo{title}{Network structure of cascading neural systems
		predicts stimulus propagation and recovery}.
	\newblock \emph{\bibinfo{journal}{Journal of Neural Engineering}}
	\textbf{\bibinfo{volume}{17}}, \bibinfo{pages}{056045}
	(\bibinfo{year}{2020}).
	
	\bibitem{wolff_intrinsic_2022}
	\bibinfo{author}{Wolff, A.} \emph{et~al.}
	\newblock \bibinfo{title}{Intrinsic neural timescales: temporal integration and
		segregation}.
	\newblock \emph{\bibinfo{journal}{Trends in Cognitive Sciences}}
	\bibinfo{pages}{S1364661321002928} (\bibinfo{year}{2022}).
	
	\bibitem{mesulam_representation_2008}
	\bibinfo{author}{Mesulam, M.}
	\newblock \bibinfo{title}{Representation, inference, and transcendent encoding
		in neurocognitive networks of the human brain}.
	\newblock \emph{\bibinfo{journal}{Annals of Neurology}}
	\textbf{\bibinfo{volume}{64}}, \bibinfo{pages}{367--378}
	(\bibinfo{year}{2008}).
	
	\bibitem{sydnor_neurodevelopment_2021}
	\bibinfo{author}{Sydnor, V.~J.} \emph{et~al.}
	\newblock \bibinfo{title}{Neurodevelopment of the association cortices:
		{Patterns}, mechanisms, and implications for psychopathology}.
	\newblock \emph{\bibinfo{journal}{Neuron}} \textbf{\bibinfo{volume}{109}},
	\bibinfo{pages}{2820--2846} (\bibinfo{year}{2021}).
	
	\bibitem{margulies_situating_2016}
	\bibinfo{author}{Margulies, D.~S.} \emph{et~al.}
	\newblock \bibinfo{title}{Situating the default-mode network along a principal
		gradient of macroscale cortical organization}.
	\newblock \emph{\bibinfo{journal}{Proceedings of the National Academy of
			Sciences}} \textbf{\bibinfo{volume}{113}}, \bibinfo{pages}{12574--12579}
	(\bibinfo{year}{2016}).
	
	\bibitem{xu_cross-species_2020}
	\bibinfo{author}{Xu, T.} \emph{et~al.}
	\newblock \bibinfo{title}{Cross-species functional alignment reveals
		evolutionary hierarchy within the connectome}.
	\newblock \emph{\bibinfo{journal}{NeuroImage}} \textbf{\bibinfo{volume}{223}},
	\bibinfo{pages}{117346} (\bibinfo{year}{2020}).
	
	\bibitem{hasson_hierarchy_2008}
	\bibinfo{author}{Hasson, U.}, \bibinfo{author}{Yang, E.},
	\bibinfo{author}{Vallines, I.}, \bibinfo{author}{Heeger, D.~J.} \&
	\bibinfo{author}{Rubin, N.}
	\newblock \bibinfo{title}{A {Hierarchy} of {Temporal} {Receptive} {Windows} in
		{Human} {Cortex}}.
	\newblock \emph{\bibinfo{journal}{Journal of Neuroscience}}
	\textbf{\bibinfo{volume}{28}}, \bibinfo{pages}{2539--2550}
	(\bibinfo{year}{2008}).
	
	\bibitem{gao_neuronal_2020}
	\bibinfo{author}{Gao, R.}, \bibinfo{author}{van~den Brink, R.~L.},
	\bibinfo{author}{Pfeffer, T.} \& \bibinfo{author}{Voytek, B.}
	\newblock \bibinfo{title}{Neuronal timescales are functionally dynamic and
		shaped by cortical microarchitecture}.
	\newblock \emph{\bibinfo{journal}{eLife}} \textbf{\bibinfo{volume}{9}},
	\bibinfo{pages}{e61277} (\bibinfo{year}{2020}).
	
	\bibitem{sydnor_intrinsic_2023}
	\bibinfo{author}{Sydnor, V.~J.} \emph{et~al.}
	\newblock \bibinfo{title}{Intrinsic activity development unfolds along a
		sensorimotor–association cortical axis in youth}.
	\newblock \emph{\bibinfo{journal}{Nature Neuroscience}}
	\textbf{\bibinfo{volume}{26}}, \bibinfo{pages}{638--649}
	(\bibinfo{year}{2023}).
	
	\bibitem{hespanha2018linear}
	\bibinfo{author}{Hespanha, J.~P.}
	\newblock \emph{\bibinfo{title}{Linear systems theory}}
	(\bibinfo{publisher}{Princeton university press}, \bibinfo{year}{2018}).
	
	\bibitem{marblestone_toward_2016}
	\bibinfo{author}{Marblestone, A.~H.}, \bibinfo{author}{Wayne, G.} \&
	\bibinfo{author}{Kording, K.~P.}
	\newblock \bibinfo{title}{Toward an {Integration} of {Deep} {Learning} and
		{Neuroscience}}.
	\newblock \emph{\bibinfo{journal}{Frontiers in Computational Neuroscience}}
	\textbf{\bibinfo{volume}{10}} (\bibinfo{year}{2016}).
	
	\bibitem{ali_predictive_2022}
	\bibinfo{author}{Ali, A.}, \bibinfo{author}{Ahmad, N.},
	\bibinfo{author}{De~Groot, E.}, \bibinfo{author}{Johannes Van~Gerven, M.~A.}
	\& \bibinfo{author}{Kietzmann, T.~C.}
	\newblock \bibinfo{title}{Predictive coding is a consequence of energy
		efficiency in recurrent neural networks}.
	\newblock \emph{\bibinfo{journal}{Patterns}} \textbf{\bibinfo{volume}{3}},
	\bibinfo{pages}{100639} (\bibinfo{year}{2022}).
	
	\bibitem{millidge_predictive_2021}
	\bibinfo{author}{Millidge, B.}, \bibinfo{author}{Seth, A.} \&
	\bibinfo{author}{Buckley, C.~L.}
	\newblock \bibinfo{title}{Predictive {Coding}: a {Theoretical} and
		{Experimental} {Review}}  (\bibinfo{year}{2021}).
	\newblock \bibinfo{note}{Publisher: arXiv Version Number: 4}.
	
	\bibitem{bastos_canonical_2012}
	\bibinfo{author}{Bastos, A.~M.} \emph{et~al.}
	\newblock \bibinfo{title}{Canonical {Microcircuits} for {Predictive} {Coding}}.
	\newblock \emph{\bibinfo{journal}{Neuron}} \textbf{\bibinfo{volume}{76}},
	\bibinfo{pages}{695--711} (\bibinfo{year}{2012}).
	
	\bibitem{singer_recurrent_2021}
	\bibinfo{author}{Singer, W.}
	\newblock \bibinfo{title}{Recurrent dynamics in the cerebral cortex:
		{Integration} of sensory evidence with stored knowledge}.
	\newblock \emph{\bibinfo{journal}{Proceedings of the National Academy of
			Sciences}} \textbf{\bibinfo{volume}{118}}, \bibinfo{pages}{e2101043118}
	(\bibinfo{year}{2021}).
	
	\bibitem{friston_computational_2022}
	\bibinfo{author}{Friston, K.}
	\newblock \bibinfo{title}{Computational psychiatry: from synapses to
		sentience}.
	\newblock \emph{\bibinfo{journal}{Molecular Psychiatry}}
	(\bibinfo{year}{2022}).
	
	\bibitem{jaeger2002tutorial}
	\bibinfo{author}{Jaeger, H.}
	\newblock \bibinfo{title}{Tutorial on training recurrent neural networks,
		covering bppt, rtrl, ekf and the" echo state network" approach}
	(\bibinfo{year}{2002}).
	
	\bibitem{tanaka2019recent}
	\bibinfo{author}{Tanaka, G.} \emph{et~al.}
	\newblock \bibinfo{title}{Recent advances in physical reservoir computing: A
		review}.
	\newblock \emph{\bibinfo{journal}{Neural Networks}}
	\textbf{\bibinfo{volume}{115}}, \bibinfo{pages}{100--123}
	(\bibinfo{year}{2019}).
	
	\bibitem{van2017advances}
	\bibinfo{author}{Van~der Sande, G.}, \bibinfo{author}{Brunner, D.} \&
	\bibinfo{author}{Soriano, M.~C.}
	\newblock \bibinfo{title}{Advances in photonic reservoir computing}.
	\newblock \emph{\bibinfo{journal}{Nanophotonics}} \textbf{\bibinfo{volume}{6}},
	\bibinfo{pages}{561--576} (\bibinfo{year}{2017}).
	
	\bibitem{soriano2014delay}
	\bibinfo{author}{Soriano, M.~C.} \emph{et~al.}
	\newblock \bibinfo{title}{Delay-based reservoir computing: noise effects in a
		combined analog and digital implementation}.
	\newblock \emph{\bibinfo{journal}{IEEE transactions on neural networks and
			learning systems}} \textbf{\bibinfo{volume}{26}}, \bibinfo{pages}{388--393}
	(\bibinfo{year}{2014}).
	
	\bibitem{suarez_learning_2021}
	\bibinfo{author}{Suárez, L.~E.}, \bibinfo{author}{Richards, B.~A.},
	\bibinfo{author}{Lajoie, G.} \& \bibinfo{author}{Misic, B.}
	\newblock \bibinfo{title}{Learning function from structure in neuromorphic
		networks}.
	\newblock \emph{\bibinfo{journal}{Nature Machine Intelligence}}
	(\bibinfo{year}{2021}).
	
	\bibitem{parkes_asymmetric_2022}
	\bibinfo{author}{Parkes, L.} \emph{et~al.}
	\newblock \bibinfo{title}{Asymmetric signaling across the hierarchy of
		cytoarchitecture within the human connectome}.
	\newblock \emph{\bibinfo{journal}{Science Advances}}
	\textbf{\bibinfo{volume}{8}}, \bibinfo{pages}{eadd2185}
	(\bibinfo{year}{2022}).
	
	\bibitem{pines_development_2023}
	\bibinfo{author}{Pines, A.} \emph{et~al.}
	\newblock \bibinfo{title}{Development of top-down cortical propagations in
		youth}.
	\newblock \emph{\bibinfo{journal}{Neuron}} \bibinfo{pages}{S0896627323000387}
	(\bibinfo{year}{2023}).
	
	\bibitem{baum_modular_2017}
	\bibinfo{author}{Baum, G.~L.} \emph{et~al.}
	\newblock \bibinfo{title}{Modular {Segregation} of {Structural} {Brain}
		{Networks} {Supports} the {Development} of {Executive} {Function} in
		{Youth}}.
	\newblock \emph{\bibinfo{journal}{Current Biology}}
	\textbf{\bibinfo{volume}{27}}, \bibinfo{pages}{1561--1572.e8}
	(\bibinfo{year}{2017}).
	
	\bibitem{shine_computational_2021}
	\bibinfo{author}{Shine, J.~M.} \emph{et~al.}
	\newblock \bibinfo{title}{Computational models link cellular mechanisms of
		neuromodulation to large-scale neural dynamics}.
	\newblock \emph{\bibinfo{journal}{Nature Neuroscience}}
	\textbf{\bibinfo{volume}{24}}, \bibinfo{pages}{765--776}
	(\bibinfo{year}{2021}).
	
	\bibitem{marder_cellular_2002}
	\bibinfo{author}{Marder, E.} \& \bibinfo{author}{Thirumalai, V.}
	\newblock \bibinfo{title}{Cellular, synaptic and network effects of
		neuromodulation}.
	\newblock \emph{\bibinfo{journal}{Neural Networks}}
	\textbf{\bibinfo{volume}{15}}, \bibinfo{pages}{479--493}
	(\bibinfo{year}{2002}).
	
	\bibitem{bucher_beyond_2011}
	\bibinfo{author}{Bucher, D.} \& \bibinfo{author}{Goaillard, J.-M.}
	\newblock \bibinfo{title}{Beyond faithful conduction: {Short}-term dynamics,
		neuromodulation, and long-term regulation of spike propagation in the axon}.
	\newblock \emph{\bibinfo{journal}{Progress in Neurobiology}}
	\textbf{\bibinfo{volume}{94}}, \bibinfo{pages}{307--346}
	(\bibinfo{year}{2011}).
	
	\bibitem{mccormick_neuromodulation_2020}
	\bibinfo{author}{McCormick, D.~A.}, \bibinfo{author}{Nestvogel, D.~B.} \&
	\bibinfo{author}{He, B.~J.}
	\newblock \bibinfo{title}{Neuromodulation of {Brain} {State} and {Behavior}}.
	\newblock \emph{\bibinfo{journal}{Annual Review of Neuroscience}}
	\textbf{\bibinfo{volume}{43}}, \bibinfo{pages}{391--415}
	(\bibinfo{year}{2020}).
	
	\bibitem{kim_brain-wide_2017}
	\bibinfo{author}{Kim, Y.} \emph{et~al.}
	\newblock \bibinfo{title}{Brain-wide {Maps} {Reveal} {Stereotyped}
		{Cell}-{Type}-{Based} {Cortical} {Architecture} and {Subcortical} {Sexual}
		{Dimorphism}}.
	\newblock \emph{\bibinfo{journal}{Cell}} \textbf{\bibinfo{volume}{171}},
	\bibinfo{pages}{456--469.e22} (\bibinfo{year}{2017}).
	
	\bibitem{burt_hierarchy_2018}
	\bibinfo{author}{Burt, J.~B.} \emph{et~al.}
	\newblock \bibinfo{title}{Hierarchy of transcriptomic specialization across
		human cortex captured by structural neuroimaging topography}.
	\newblock \emph{\bibinfo{journal}{Nature Neuroscience}}
	\textbf{\bibinfo{volume}{21}}, \bibinfo{pages}{1251--1259}
	(\bibinfo{year}{2018}).
	
	\bibitem{anderson_transcriptional_2020}
	\bibinfo{author}{Anderson, K.~M.} \emph{et~al.}
	\newblock \bibinfo{title}{Transcriptional and imaging-genetic association of
		cortical interneurons, brain function, and schizophrenia risk}.
	\newblock \emph{\bibinfo{journal}{Nature Communications}}
	\textbf{\bibinfo{volume}{11}}, \bibinfo{pages}{2889} (\bibinfo{year}{2020}).
	
	\bibitem{deco_how_2014}
	\bibinfo{author}{Deco, G.} \emph{et~al.}
	\newblock \bibinfo{title}{How {Local} {Excitation}-{Inhibition} {Ratio}
		{Impacts} the {Whole} {Brain} {Dynamics}}.
	\newblock \emph{\bibinfo{journal}{Journal of Neuroscience}}
	\textbf{\bibinfo{volume}{34}}, \bibinfo{pages}{7886--7898}
	(\bibinfo{year}{2014}).
	
	\bibitem{deco_dynamical_2021}
	\bibinfo{author}{Deco, G.} \emph{et~al.}
	\newblock \bibinfo{title}{Dynamical consequences of regional heterogeneity in
		the brain’s transcriptional landscape}.
	\newblock \emph{\bibinfo{journal}{Science Advances}}
	\textbf{\bibinfo{volume}{7}}, \bibinfo{pages}{eabf4752}
	(\bibinfo{year}{2021}).
	
	\bibitem{gao_inferring_2017}
	\bibinfo{author}{Gao, R.}, \bibinfo{author}{Peterson, E.~J.} \&
	\bibinfo{author}{Voytek, B.}
	\newblock \bibinfo{title}{Inferring synaptic excitation/inhibition balance from
		field potentials}.
	\newblock \emph{\bibinfo{journal}{NeuroImage}} \textbf{\bibinfo{volume}{158}},
	\bibinfo{pages}{70--78} (\bibinfo{year}{2017}).
	
	\bibitem{zhang_ei_2023}
	\bibinfo{author}{Zhang, S.} \emph{et~al.}
	\newblock \bibinfo{title}{In-vivo whole-cortex estimation of
		excitation-inhibition ratio indexes cortical maturation and cognitive ability
		in youth}.
	\newblock \bibinfo{type}{preprint}, \bibinfo{institution}{bioRxiv}
	(\bibinfo{year}{2023}).
	
	\bibitem{larsen_developmental_2022}
	\bibinfo{author}{Larsen, B.} \emph{et~al.}
	\newblock \bibinfo{title}{A developmental reduction of the
		excitation:inhibition ratio in association cortex during adolescence}.
	\newblock \emph{\bibinfo{journal}{Science Advances}}
	\textbf{\bibinfo{volume}{8}}, \bibinfo{pages}{eabj8750}
	(\bibinfo{year}{2022}).
	
	\bibitem{rocchi_increased_2022}
	\bibinfo{author}{Rocchi, F.} \emph{et~al.}
	\newblock \bibinfo{title}{Increased {fMRI} connectivity upon chemogenetic
		inhibition of the mouse prefrontal cortex}.
	\newblock \emph{\bibinfo{journal}{Nature Communications}}
	\textbf{\bibinfo{volume}{13}}, \bibinfo{pages}{1056} (\bibinfo{year}{2022}).
	
	\bibitem{markicevic_cortical_2020}
	\bibinfo{author}{Markicevic, M.} \emph{et~al.}
	\newblock \bibinfo{title}{Cortical {Excitation}:{Inhibition} {Imbalance}
		{Causes} {Abnormal} {Brain} {Network} {Dynamics} as {Observed} in
		{Neurodevelopmental} {Disorders}}.
	\newblock \emph{\bibinfo{journal}{Cerebral Cortex}}
	\textbf{\bibinfo{volume}{30}}, \bibinfo{pages}{4922--4937}
	(\bibinfo{year}{2020}).
	
	\bibitem{shine_modulation_2018}
	\bibinfo{author}{Shine, J.~M.}, \bibinfo{author}{Aburn, M.~J.},
	\bibinfo{author}{Breakspear, M.} \& \bibinfo{author}{Poldrack, R.~A.}
	\newblock \bibinfo{title}{The modulation of neural gain facilitates a
		transition between functional segregation and integration in the brain}.
	\newblock \emph{\bibinfo{journal}{eLife}} \textbf{\bibinfo{volume}{7}},
	\bibinfo{pages}{e31130} (\bibinfo{year}{2018}).
	
	\bibitem{shine_human_2019}
	\bibinfo{author}{Shine, J.~M.} \emph{et~al.}
	\newblock \bibinfo{title}{Human cognition involves the dynamic integration of
		neural activity and neuromodulatory systems}.
	\newblock \emph{\bibinfo{journal}{Nature Neuroscience}}
	\textbf{\bibinfo{volume}{22}}, \bibinfo{pages}{289--296}
	(\bibinfo{year}{2019}).
	
	\bibitem{bertolero_modular_2015}
	\bibinfo{author}{Bertolero, M.~A.}, \bibinfo{author}{Yeo, B. T.~T.} \&
	\bibinfo{author}{D’Esposito, M.}
	\newblock \bibinfo{title}{The modular and integrative functional architecture
		of the human brain}.
	\newblock \emph{\bibinfo{journal}{Proceedings of the National Academy of
			Sciences}} \textbf{\bibinfo{volume}{112}} (\bibinfo{year}{2015}).
	
	\bibitem{vazquez-rodriguez_gradients_2019}
	\bibinfo{author}{Vázquez-Rodríguez, B.} \emph{et~al.}
	\newblock \bibinfo{title}{Gradients of structure–function tethering across
		neocortex}.
	\newblock \emph{\bibinfo{journal}{Proceedings of the National Academy of
			Sciences}} \textbf{\bibinfo{volume}{116}}, \bibinfo{pages}{21219--21227}
	(\bibinfo{year}{2019}).
	
	\bibitem{preti_decoupling_2019}
	\bibinfo{author}{Preti, M.~G.} \& \bibinfo{author}{Van De~Ville, D.}
	\newblock \bibinfo{title}{Decoupling of brain function from structure reveals
		regional behavioral specialization in humans}.
	\newblock \emph{\bibinfo{journal}{Nature Communications}}
	\textbf{\bibinfo{volume}{10}}, \bibinfo{pages}{4747} (\bibinfo{year}{2019}).
	
	\bibitem{misic_network-level_2016}
	\bibinfo{author}{Mišić, B.} \emph{et~al.}
	\newblock \bibinfo{title}{Network-{Level} {Structure}-{Function}
		{Relationships} in {Human} {Neocortex}}.
	\newblock \emph{\bibinfo{journal}{Cerebral Cortex}}
	\textbf{\bibinfo{volume}{26}}, \bibinfo{pages}{3285--3296}
	(\bibinfo{year}{2016}).
	
	\bibitem{zamani_esfahlani_local_2022}
	\bibinfo{author}{Zamani~Esfahlani, F.}, \bibinfo{author}{Faskowitz, J.},
	\bibinfo{author}{Slack, J.}, \bibinfo{author}{Mišić, B.} \&
	\bibinfo{author}{Betzel, R.~F.}
	\newblock \bibinfo{title}{Local structure-function relationships in human brain
		networks across the lifespan}.
	\newblock \emph{\bibinfo{journal}{Nature Communications}}
	\textbf{\bibinfo{volume}{13}}, \bibinfo{pages}{2053} (\bibinfo{year}{2022}).
	
	\bibitem{samuels_functional_2008}
	\bibinfo{author}{Samuels, E.} \& \bibinfo{author}{Szabadi, E.}
	\newblock \bibinfo{title}{Functional {Neuroanatomy} of the {Noradrenergic}
		{Locus} {Coeruleus}: {Its} {Roles} in the {Regulation} of {Arousal} and
		{Autonomic} {Function} {Part} {II}: {Physiological} and {Pharmacological}
		{Manipulations} and {Pathological} {Alterations} of {Locus} {Coeruleus}
		{Activity} in {Humans}}.
	\newblock \emph{\bibinfo{journal}{Current Neuropharmacology}}
	\textbf{\bibinfo{volume}{6}}, \bibinfo{pages}{254--285}
	(\bibinfo{year}{2008}).
	
	\bibitem{wainstein_gain_2023}
	\bibinfo{author}{Wainstein, G.} \emph{et~al.}
	\newblock \bibinfo{title}{Gain neuromodulation mediates perceptual switches:
		evidence from pupillometry, {fMRI}, and {RNN} {Modelling}}.
	\newblock \bibinfo{type}{preprint}, \bibinfo{institution}{Research Square}
	(\bibinfo{year}{2023}).
	
	\bibitem{joshi_pupil_2020}
	\bibinfo{author}{Joshi, S.} \& \bibinfo{author}{Gold, J.~I.}
	\newblock \bibinfo{title}{Pupil {Size} as a {Window} on {Neural} {Substrates}
		of {Cognition}}.
	\newblock \emph{\bibinfo{journal}{Trends in Cognitive Sciences}}
	\textbf{\bibinfo{volume}{24}}, \bibinfo{pages}{466--480}
	(\bibinfo{year}{2020}).
	
	\bibitem{singleton_receptor-informed_2022}
	\bibinfo{author}{Singleton, S.~P.} \emph{et~al.}
	\newblock \bibinfo{title}{Receptor-informed network control theory links {LSD}
		and psilocybin to a flattening of the brain’s control energy landscape}.
	\newblock \emph{\bibinfo{journal}{Nature Communications}}
	\textbf{\bibinfo{volume}{13}}, \bibinfo{pages}{5812} (\bibinfo{year}{2022}).
	
	\bibitem{luppi_transitions_2023}
	\bibinfo{author}{Luppi, A.~I.} \emph{et~al.}
	\newblock \bibinfo{title}{Transitions between cognitive topographies:
		contributions of network structure, neuromodulation, and disease}.
	\newblock \bibinfo{type}{preprint}, \bibinfo{institution}{bioRxiv}
	(\bibinfo{year}{2023}).
	
	\bibitem{aitken2023neural}
	\bibinfo{author}{Aitken, K.} \& \bibinfo{author}{Mihalas, S.}
	\newblock \bibinfo{title}{Neural population dynamics of computing with synaptic
		modulations}.
	\newblock \emph{\bibinfo{journal}{Elife}} \textbf{\bibinfo{volume}{12}},
	\bibinfo{pages}{e83035} (\bibinfo{year}{2023}).
	
	\bibitem{jonas2017could}
	\bibinfo{author}{Jonas, E.} \& \bibinfo{author}{Kording, K.~P.}
	\newblock \bibinfo{title}{Could a neuroscientist understand a microprocessor?}
	\newblock \emph{\bibinfo{journal}{PLoS computational biology}}
	\textbf{\bibinfo{volume}{13}}, \bibinfo{pages}{e1005268}
	(\bibinfo{year}{2017}).
	
	\bibitem{torii2016asic}
	\bibinfo{author}{Torii, N.} \emph{et~al.}
	\newblock \bibinfo{title}{Asic implementation of random number generators using
		sr latches and its evaluation}.
	\newblock \emph{\bibinfo{journal}{EURASIP Journal on Information Security}}
	\textbf{\bibinfo{volume}{2016}}, \bibinfo{pages}{1--12}
	(\bibinfo{year}{2016}).
	
	\bibitem{samsonovich1997path}
	\bibinfo{author}{Samsonovich, A.} \& \bibinfo{author}{McNaughton, B.~L.}
	\newblock \bibinfo{title}{Path integration and cognitive mapping in a
		continuous attractor neural network model}.
	\newblock \emph{\bibinfo{journal}{Journal of Neuroscience}}
	\textbf{\bibinfo{volume}{17}}, \bibinfo{pages}{5900--5920}
	(\bibinfo{year}{1997}).
	
	\bibitem{fung2010moving}
	\bibinfo{author}{Fung, C.~A.}, \bibinfo{author}{Wong, K.~M.} \&
	\bibinfo{author}{Wu, S.}
	\newblock \bibinfo{title}{A moving bump in a continuous manifold: a
		comprehensive study of the tracking dynamics of continuous attractor neural
		networks}.
	\newblock \emph{\bibinfo{journal}{Neural Computation}}
	\textbf{\bibinfo{volume}{22}}, \bibinfo{pages}{752--792}
	(\bibinfo{year}{2010}).
	
	\bibitem{wimmer2014bump}
	\bibinfo{author}{Wimmer, K.}, \bibinfo{author}{Nykamp, D.~Q.},
	\bibinfo{author}{Constantinidis, C.} \& \bibinfo{author}{Compte, A.}
	\newblock \bibinfo{title}{Bump attractor dynamics in prefrontal cortex explains
		behavioral precision in spatial working memory}.
	\newblock \emph{\bibinfo{journal}{Nature neuroscience}}
	\textbf{\bibinfo{volume}{17}}, \bibinfo{pages}{431--439}
	(\bibinfo{year}{2014}).
	
	\bibitem{hopfield1982neural}
	\bibinfo{author}{Hopfield, J.~J.}
	\newblock \bibinfo{title}{Neural networks and physical systems with emergent
		collective computational abilities.}
	\newblock \emph{\bibinfo{journal}{Proceedings of the national academy of
			sciences}} \textbf{\bibinfo{volume}{79}}, \bibinfo{pages}{2554--2558}
	(\bibinfo{year}{1982}).
	
	\bibitem{strogatz2018nonlinear}
	\bibinfo{author}{Strogatz, S.~H.}
	\newblock \emph{\bibinfo{title}{Nonlinear dynamics and chaos with student
			solutions manual: With applications to physics, biology, chemistry, and
			engineering}} (\bibinfo{publisher}{CRC press}, \bibinfo{year}{2018}).
	
	\bibitem{ramsauer2020hopfield}
	\bibinfo{author}{Ramsauer, H.} \emph{et~al.}
	\newblock \bibinfo{title}{Hopfield networks is all you need}.
	\newblock \emph{\bibinfo{journal}{arXiv preprint arXiv:2008.02217}}
	(\bibinfo{year}{2020}).
	
	\bibitem{storkey1997increasing}
	\bibinfo{author}{Storkey, A.}
	\newblock \bibinfo{title}{Increasing the capacity of a hopfield network without
		sacrificing functionality}.
	\newblock In \emph{\bibinfo{booktitle}{Artificial Neural Networks—ICANN'97:
			7th International Conference Lausanne, Switzerland, October 8--10, 1997
			Proceeedings 7}}, \bibinfo{pages}{451--456}
	(\bibinfo{organization}{Springer}, \bibinfo{year}{1997}).
	
	\bibitem{smith2022learning}
	\bibinfo{author}{Smith, L.~M.}, \bibinfo{author}{Kim, J.~Z.},
	\bibinfo{author}{Lu, Z.} \& \bibinfo{author}{Bassett, D.~S.}
	\newblock \bibinfo{title}{Learning continuous chaotic attractors with a
		reservoir computer}.
	\newblock \emph{\bibinfo{journal}{Chaos: An Interdisciplinary Journal of
			Nonlinear Science}} \textbf{\bibinfo{volume}{32}} (\bibinfo{year}{2022}).
	
	\bibitem{nichols2002middle}
	\bibinfo{author}{Nichols, M.~J.} \& \bibinfo{author}{Newsome, W.~T.}
	\newblock \bibinfo{title}{Middle temporal visual area microstimulation
		influences veridical judgments of motion direction}.
	\newblock \emph{\bibinfo{journal}{Journal of Neuroscience}}
	\textbf{\bibinfo{volume}{22}}, \bibinfo{pages}{9530--9540}
	(\bibinfo{year}{2002}).
	
	\bibitem{jazayeri2021interpreting}
	\bibinfo{author}{Jazayeri, M.} \& \bibinfo{author}{Ostojic, S.}
	\newblock \bibinfo{title}{Interpreting neural computations by examining
		intrinsic and embedding dimensionality of neural activity}.
	\newblock \emph{\bibinfo{journal}{Current opinion in neurobiology}}
	\textbf{\bibinfo{volume}{70}}, \bibinfo{pages}{113--120}
	(\bibinfo{year}{2021}).
	
	\bibitem{howarth_updated_2012}
	\bibinfo{author}{Howarth, C.}, \bibinfo{author}{Gleeson, P.} \&
	\bibinfo{author}{Attwell, D.}
	\newblock \bibinfo{title}{Updated {Energy} {Budgets} for {Neural} {Computation}
		in the {Neocortex} and {Cerebellum}}.
	\newblock \emph{\bibinfo{journal}{Journal of Cerebral Blood Flow \&
			Metabolism}} \textbf{\bibinfo{volume}{32}}, \bibinfo{pages}{1222--1232}
	(\bibinfo{year}{2012}).
	
	\bibitem{giesl2015review}
	\bibinfo{author}{Giesl, P.} \& \bibinfo{author}{Hafstein, S.}
	\newblock \bibinfo{title}{Review on computational methods for lyapunov
		functions}.
	\newblock \emph{\bibinfo{journal}{Discrete and Continuous Dynamical Systems-B}}
	\textbf{\bibinfo{volume}{20}}, \bibinfo{pages}{2291--2331}
	(\bibinfo{year}{2015}).
	
	\bibitem{bellman1962vector}
	\bibinfo{author}{Bellman, R.}
	\newblock \bibinfo{title}{Vector lyapunov functions}.
	\newblock \emph{\bibinfo{journal}{Journal of the Society for Industrial and
			Applied Mathematics, Series A: Control}} \textbf{\bibinfo{volume}{1}},
	\bibinfo{pages}{32--34} (\bibinfo{year}{1962}).
	
	\bibitem{wolf1985determining}
	\bibinfo{author}{Wolf, A.}, \bibinfo{author}{Swift, J.~B.},
	\bibinfo{author}{Swinney, H.~L.} \& \bibinfo{author}{Vastano, J.~A.}
	\newblock \bibinfo{title}{Determining lyapunov exponents from a time series}.
	\newblock \emph{\bibinfo{journal}{Physica D: nonlinear phenomena}}
	\textbf{\bibinfo{volume}{16}}, \bibinfo{pages}{285--317}
	(\bibinfo{year}{1985}).
	
	\bibitem{petanjek_extraordinary_2011}
	\bibinfo{author}{Petanjek, Z.} \emph{et~al.}
	\newblock \bibinfo{title}{Extraordinary neoteny of synaptic spines in the human
		prefrontal cortex}.
	\newblock \emph{\bibinfo{journal}{Proceedings of the National Academy of
			Sciences}} \textbf{\bibinfo{volume}{108}}, \bibinfo{pages}{13281--13286}
	(\bibinfo{year}{2011}).
	
	\bibitem{averbeck_pruning_2022}
	\bibinfo{author}{Averbeck, B.~B.}
	\newblock \bibinfo{title}{Pruning recurrent neural networks replicates
		adolescent changes in working memory and reinforcement learning}.
	\newblock \emph{\bibinfo{journal}{Proceedings of the National Academy of
			Sciences}} \textbf{\bibinfo{volume}{119}}, \bibinfo{pages}{e2121331119}
	(\bibinfo{year}{2022}).
	
	\bibitem{moler1967iterative}
	\bibinfo{author}{Moler, C.~B.}
	\newblock \bibinfo{title}{Iterative refinement in floating point}.
	\newblock \emph{\bibinfo{journal}{Journal of the ACM (JACM)}}
	\textbf{\bibinfo{volume}{14}}, \bibinfo{pages}{316--321}
	(\bibinfo{year}{1967}).
	
	\bibitem{ypma1995historical}
	\bibinfo{author}{Ypma, T.~J.}
	\newblock \bibinfo{title}{Historical development of the newton--raphson
		method}.
	\newblock \emph{\bibinfo{journal}{SIAM review}} \textbf{\bibinfo{volume}{37}},
	\bibinfo{pages}{531--551} (\bibinfo{year}{1995}).
	
	\bibitem{aragon2020douglas}
	\bibinfo{author}{Arag{\'o}n~Artacho, F.~J.}, \bibinfo{author}{Campoy, R.} \&
	\bibinfo{author}{Tam, M.~K.}
	\newblock \bibinfo{title}{The douglas--rachford algorithm for convex and
		nonconvex feasibility problems}.
	\newblock \emph{\bibinfo{journal}{Mathematical Methods of Operations Research}}
	\textbf{\bibinfo{volume}{91}}, \bibinfo{pages}{201--240}
	(\bibinfo{year}{2020}).
	
	\bibitem{flesch2022orthogonal}
	\bibinfo{author}{Flesch, T.}, \bibinfo{author}{Juechems, K.},
	\bibinfo{author}{Dumbalska, T.}, \bibinfo{author}{Saxe, A.} \&
	\bibinfo{author}{Summerfield, C.}
	\newblock \bibinfo{title}{Orthogonal representations for robust
		context-dependent task performance in brains and neural networks}.
	\newblock \emph{\bibinfo{journal}{Neuron}} \textbf{\bibinfo{volume}{110}},
	\bibinfo{pages}{1258--1270} (\bibinfo{year}{2022}).
	
	\bibitem{gillespie2021hippocampal}
	\bibinfo{author}{Gillespie, A.~K.} \emph{et~al.}
	\newblock \bibinfo{title}{Hippocampal replay reflects specific past experiences
		rather than a plan for subsequent choice}.
	\newblock \emph{\bibinfo{journal}{Neuron}} \textbf{\bibinfo{volume}{109}},
	\bibinfo{pages}{3149--3163} (\bibinfo{year}{2021}).
	
	\bibitem{rajalingham2022dynamic}
	\bibinfo{author}{Rajalingham, R.}, \bibinfo{author}{Sohn, H.} \&
	\bibinfo{author}{Jazayeri, M.}
	\newblock \bibinfo{title}{Dynamic tracking of objects in the macaque
		dorsomedial frontal cortex}.
	\newblock \emph{\bibinfo{journal}{bioRxiv}} \bibinfo{pages}{2022--06}
	(\bibinfo{year}{2022}).
	
	\bibitem{rajalingham2022recurrent}
	\bibinfo{author}{Rajalingham, R.}, \bibinfo{author}{Piccato, A.} \&
	\bibinfo{author}{Jazayeri, M.}
	\newblock \bibinfo{title}{Recurrent neural networks with explicit
		representation of dynamic latent variables can mimic behavioral patterns in a
		physical inference task}.
	\newblock \emph{\bibinfo{journal}{Nature Communications}}
	\textbf{\bibinfo{volume}{13}}, \bibinfo{pages}{5865} (\bibinfo{year}{2022}).
	
	\bibitem{ercsey2011optimization}
	\bibinfo{author}{Ercsey-Ravasz, M.} \& \bibinfo{author}{Toroczkai, Z.}
	\newblock \bibinfo{title}{Optimization hardness as transient chaos in an analog
		approach to constraint satisfaction}.
	\newblock \emph{\bibinfo{journal}{Nature Physics}}
	\textbf{\bibinfo{volume}{7}}, \bibinfo{pages}{966--970}
	(\bibinfo{year}{2011}).
	
	\bibitem{molnar2012continuous}
	\bibinfo{author}{Molnar, B.}, \bibinfo{author}{Toroczkai, Z.} \&
	\bibinfo{author}{Ercsey-Ravasz, M.}
	\newblock \bibinfo{title}{Continuous-time neural networks without local traps
		for solving boolean satisfiability}.
	\newblock In \emph{\bibinfo{booktitle}{2012 13th International Workshop on
			Cellular Nanoscale Networks and their Applications}}, \bibinfo{pages}{1--6}
	(\bibinfo{organization}{IEEE}, \bibinfo{year}{2012}).
	
	\bibitem{yamashita2019bounded}
	\bibinfo{author}{Yamashita, H.}, \bibinfo{author}{Suzuki, H.},
	\bibinfo{author}{Toroczkai, Z.} \& \bibinfo{author}{Aihara, K.}
	\newblock \bibinfo{title}{Bounded continuous-time satisfiability solver}.
	\newblock In \emph{\bibinfo{booktitle}{International Symposium on Nonlinear
			Theory and its Applications (NOLTA2019)}} (\bibinfo{year}{2019}).
	
	\bibitem{li2019continuous}
	\bibinfo{author}{Li, C.} \& \bibinfo{author}{MacLennan, B.~J.}
	\newblock \bibinfo{title}{Continuous-time systems for solving 0-1 integer
		linear programming feasibility problems}.
	\newblock \emph{\bibinfo{journal}{arXiv preprint arXiv:1905.04612}}
	(\bibinfo{year}{2019}).
	
	\bibitem{ding2010high}
	\bibinfo{author}{Ding, Y.}, \bibinfo{author}{Li, Y.}, \bibinfo{author}{Xiao,
		M.}, \bibinfo{author}{Wang, Q.} \& \bibinfo{author}{Li, D.}
	\newblock \bibinfo{title}{A high order neural network to solve n-queens
		problem}.
	\newblock In \emph{\bibinfo{booktitle}{The 2010 International Joint Conference
			on Neural Networks (IJCNN)}}, \bibinfo{pages}{1--6}
	(\bibinfo{organization}{IEEE}, \bibinfo{year}{2010}).
	
	\bibitem{mastrogiuseppe2018linking}
	\bibinfo{author}{Mastrogiuseppe, F.} \& \bibinfo{author}{Ostojic, S.}
	\newblock \bibinfo{title}{Linking connectivity, dynamics, and computations in
		low-rank recurrent neural networks}.
	\newblock \emph{\bibinfo{journal}{Neuron}} \textbf{\bibinfo{volume}{99}},
	\bibinfo{pages}{609--623} (\bibinfo{year}{2018}).
	
	\bibitem{eliasmith2003neural}
	\bibinfo{author}{Eliasmith, C.} \& \bibinfo{author}{Anderson, C.~H.}
	\newblock \emph{\bibinfo{title}{Neural engineering: Computation,
			representation, and dynamics in neurobiological systems}}
	(\bibinfo{publisher}{MIT press}, \bibinfo{year}{2003}).
	
	\bibitem{dewolf2020nengo}
	\bibinfo{author}{DeWolf, T.}, \bibinfo{author}{Jaworski, P.} \&
	\bibinfo{author}{Eliasmith, C.}
	\newblock \bibinfo{title}{Nengo and low-power ai hardware for robust, embedded
		neurorobotics}.
	\newblock \emph{\bibinfo{journal}{Frontiers in Neurorobotics}}
	\textbf{\bibinfo{volume}{14}}, \bibinfo{pages}{568359}
	(\bibinfo{year}{2020}).
	
	\bibitem{graves2016hybrid}
	\bibinfo{author}{Graves, A.} \emph{et~al.}
	\newblock \bibinfo{title}{Hybrid computing using a neural network with dynamic
		external memory}.
	\newblock \emph{\bibinfo{journal}{Nature}} \textbf{\bibinfo{volume}{538}},
	\bibinfo{pages}{471--476} (\bibinfo{year}{2016}).
	
	\bibitem{tkavcik2016neural}
	\bibinfo{author}{Tka{\v{c}}{\'\i}k, J.} \& \bibinfo{author}{Kord{\'\i}k, P.}
	\newblock \bibinfo{title}{Neural turing machine for sequential learning of
		human mobility patterns}.
	\newblock In \emph{\bibinfo{booktitle}{2016 International joint conference on
			neural networks (IJCNN)}}, \bibinfo{pages}{2790--2797}
	(\bibinfo{organization}{IEEE}, \bibinfo{year}{2016}).
	
	\bibitem{fawzi2022discovering}
	\bibinfo{author}{Fawzi, A.} \emph{et~al.}
	\newblock \bibinfo{title}{Discovering faster matrix multiplication algorithms
		with reinforcement learning}.
	\newblock \emph{\bibinfo{journal}{Nature}} \textbf{\bibinfo{volume}{610}},
	\bibinfo{pages}{47--53} (\bibinfo{year}{2022}).
	
	\bibitem{kim2021teaching}
	\bibinfo{author}{Kim, J.~Z.}, \bibinfo{author}{Lu, Z.},
	\bibinfo{author}{Nozari, E.}, \bibinfo{author}{Pappas, G.~J.} \&
	\bibinfo{author}{Bassett, D.~S.}
	\newblock \bibinfo{title}{Teaching recurrent neural networks to infer global
		temporal structure from local examples}.
	\newblock \emph{\bibinfo{journal}{Nature Machine Intelligence}}
	\textbf{\bibinfo{volume}{3}}, \bibinfo{pages}{316--323}
	(\bibinfo{year}{2021}).
	
	\bibitem{valente2022probing}
	\bibinfo{author}{Valente, A.}, \bibinfo{author}{Ostojic, S.} \&
	\bibinfo{author}{Pillow, J.~W.}
	\newblock \bibinfo{title}{Probing the relationship between latent linear
		dynamical systems and low-rank recurrent neural network models}.
	\newblock \emph{\bibinfo{journal}{Neural computation}}
	\textbf{\bibinfo{volume}{34}}, \bibinfo{pages}{1871--1892}
	(\bibinfo{year}{2022}).
	
	\bibitem{rajan2016recurrent}
	\bibinfo{author}{Rajan, K.}, \bibinfo{author}{Harvey, C.~D.} \&
	\bibinfo{author}{Tank, D.~W.}
	\newblock \bibinfo{title}{Recurrent network models of sequence generation and
		memory}.
	\newblock \emph{\bibinfo{journal}{Neuron}} \textbf{\bibinfo{volume}{90}},
	\bibinfo{pages}{128--142} (\bibinfo{year}{2016}).
	
	\bibitem{kepple2022curriculum}
	\bibinfo{author}{Kepple, D.}, \bibinfo{author}{Engelken, R.} \&
	\bibinfo{author}{Rajan, K.}
	\newblock \bibinfo{title}{Curriculum learning as a tool to uncover learning
		principles in the brain}.
	\newblock In \emph{\bibinfo{booktitle}{International Conference on Learning
			Representations}} (\bibinfo{year}{2022}).
	
	\bibitem{larsen_critical_2023}
	\bibinfo{author}{Larsen, B.}, \bibinfo{author}{Sydnor, V.~J.},
	\bibinfo{author}{Keller, A.~S.}, \bibinfo{author}{Yeo, B.~T.} \&
	\bibinfo{author}{Satterthwaite, T.~D.}
	\newblock \bibinfo{title}{A critical period plasticity framework for the
		sensorimotor–association axis of cortical neurodevelopment}.
	\newblock \emph{\bibinfo{journal}{Trends in Neurosciences}}
	\bibinfo{pages}{S0166223623001674} (\bibinfo{year}{2023}).
	
	\bibitem{kim_biased_2019}
	\bibinfo{author}{Kim, N.~Y.} \& \bibinfo{author}{Kastner, S.}
	\newblock \bibinfo{title}{A biased competition theory for the developmental
		cognitive neuroscience of visuo-spatial attention}.
	\newblock \emph{\bibinfo{journal}{Current Opinion in Psychology}}
	\textbf{\bibinfo{volume}{29}}, \bibinfo{pages}{219--228}
	(\bibinfo{year}{2019}).
	
	\bibitem{tervo-clemmens_canonical_2022}
	\bibinfo{author}{Tervo-Clemmens, B.} \emph{et~al.}
	\newblock \bibinfo{title}{A {Canonical} {Trajectory} of {Executive} {Function}
		{Maturation} {During} the {Transition} from {Adolescence} to {Adulthood}}.
	\newblock \bibinfo{type}{preprint}, \bibinfo{institution}{PsyArXiv}
	(\bibinfo{year}{2022}).
	
	\bibitem{larsen_adolescence_2018}
	\bibinfo{author}{Larsen, B.} \& \bibinfo{author}{Luna, B.}
	\newblock \bibinfo{title}{Adolescence as a neurobiological critical period for
		the development of higher-order cognition}.
	\newblock \emph{\bibinfo{journal}{Neuroscience \& Biobehavioral Reviews}}
	\textbf{\bibinfo{volume}{94}}, \bibinfo{pages}{179--195}
	(\bibinfo{year}{2018}).
	
	\bibitem{garcia-cabezas_structural_2019}
	\bibinfo{author}{Garc{\'\i}a-Cabezas, M.~{\'A}.}, \bibinfo{author}{Zikopoulos,
		B.} \& \bibinfo{author}{Barbas, H.}
	\newblock \bibinfo{title}{The structural model: a theory linking connections,
		plasticity, pathology, development and evolution of the cerebral cortex}.
	\newblock \emph{\bibinfo{journal}{Brain Structure and Function}}
	\textbf{\bibinfo{volume}{224}}, \bibinfo{pages}{985--1008}
	(\bibinfo{year}{2019}).
	
	\bibitem{garcia-cabezas_protocol_2020}
	\bibinfo{author}{Garc{\'\i}a-Cabezas, M.~{\'A}.}, \bibinfo{author}{Hacker,
		J.~L.} \& \bibinfo{author}{Zikopoulos, B.}
	\newblock \bibinfo{title}{A protocol for cortical type analysis of the human
		neocortex applied on histological samples, the atlas of von economo and
		koskinas, and magnetic resonance imaging}.
	\newblock \emph{\bibinfo{journal}{Frontiers in Neuroanatomy}}
	\textbf{\bibinfo{volume}{14}}, \bibinfo{pages}{576015}
	(\bibinfo{year}{2020}).
	
	\bibitem{paquola_microstructural_2019}
	\bibinfo{author}{Paquola, C.} \emph{et~al.}
	\newblock \bibinfo{title}{Microstructural and functional gradients are
		increasingly dissociated in transmodal cortices}.
	\newblock \emph{\bibinfo{journal}{PLOS Biology}} \textbf{\bibinfo{volume}{17}},
	\bibinfo{pages}{e3000284} (\bibinfo{year}{2019}).
	
	\bibitem{barbas_general_2015}
	\bibinfo{author}{Barbas, H.}
	\newblock \bibinfo{title}{General {Cortical} and {Special} {Prefrontal}
		{Connections}: {Principles} from {Structure} to {Function}}.
	\newblock \emph{\bibinfo{journal}{Annual Review of Neuroscience}}
	\textbf{\bibinfo{volume}{38}}, \bibinfo{pages}{269--289}
	(\bibinfo{year}{2015}).
	
	\bibitem{markov_anatomy_2014}
	\bibinfo{author}{Markov, N.~T.} \emph{et~al.}
	\newblock \bibinfo{title}{Anatomy of hierarchy: {Feedforward} and feedback
		pathways in macaque visual cortex}.
	\newblock \emph{\bibinfo{journal}{Journal of Comparative Neurology}}
	\textbf{\bibinfo{volume}{522}}, \bibinfo{pages}{225--259}
	(\bibinfo{year}{2014}).
	
	\bibitem{beul_towards_2015}
	\bibinfo{author}{Beul, S.~F.}
	\newblock \bibinfo{title}{Towards a "canonical" agranular cortical
		microcircuit}.
	\newblock \emph{\bibinfo{journal}{Frontiers in Neuroanatomy}}
	\bibinfo{pages}{8} (\bibinfo{year}{2015}).
	
	\bibitem{huttenlocher_synaptogenesis_1982}
	\bibinfo{author}{Huttenlocher, P.~R.}, \bibinfo{author}{de~Courten, C.},
	\bibinfo{author}{Garey, L.~J.} \& \bibinfo{author}{Van~der Loos, H.}
	\newblock \bibinfo{title}{Synaptogenesis in human visual cortex — evidence
		for synapse elimination during normal development}.
	\newblock \emph{\bibinfo{journal}{Neuroscience Letters}}
	\textbf{\bibinfo{volume}{33}}, \bibinfo{pages}{247--252}
	(\bibinfo{year}{1982}).
	
	\bibitem{peter_r_synaptic_1979}
	\bibinfo{author}{Peter~R., H.}
	\newblock \bibinfo{title}{Synaptic density in human frontal cortex —
		{Developmental} changes and effects of aging}.
	\newblock \emph{\bibinfo{journal}{Brain Research}}
	\textbf{\bibinfo{volume}{163}}, \bibinfo{pages}{195--205}
	(\bibinfo{year}{1979}).
	
	\bibitem{semple_brain_2013}
	\bibinfo{author}{Semple, B.~D.}, \bibinfo{author}{Blomgren, K.},
	\bibinfo{author}{Gimlin, K.}, \bibinfo{author}{Ferriero, D.~M.} \&
	\bibinfo{author}{Noble-Haeusslein, L.~J.}
	\newblock \bibinfo{title}{Brain development in rodents and humans:
		{Identifying} benchmarks of maturation and vulnerability to injury across
		species}.
	\newblock \emph{\bibinfo{journal}{Progress in Neurobiology}}
	\textbf{\bibinfo{volume}{106-107}}, \bibinfo{pages}{1--16}
	(\bibinfo{year}{2013}).
	
	\bibitem{buckner_evolution_2013}
	\bibinfo{author}{Buckner, R.~L.} \& \bibinfo{author}{Krienen, F.~M.}
	\newblock \bibinfo{title}{The evolution of distributed association networks in
		the human brain}.
	\newblock \emph{\bibinfo{journal}{Trends in Cognitive Sciences}}
	\textbf{\bibinfo{volume}{17}}, \bibinfo{pages}{648--665}
	(\bibinfo{year}{2013}).
	
	\bibitem{achterberg_spatially_2023}
	\bibinfo{author}{Achterberg, J.}, \bibinfo{author}{Akarca, D.},
	\bibinfo{author}{Strouse, D.~J.}, \bibinfo{author}{Duncan, J.} \&
	\bibinfo{author}{Astle, D.~E.}
	\newblock \bibinfo{title}{Spatially embedded recurrent neural networks reveal
		widespread links between structural and functional neuroscience findings}.
	\newblock \emph{\bibinfo{journal}{Nature Machine Intelligence}}
	(\bibinfo{year}{2023}).
	\newblock \urlprefix\url{https://www.nature.com/articles/s42256-023-00748-9}.
	
	\bibitem{sporns_modular_2016}
	\bibinfo{author}{Sporns, O.} \& \bibinfo{author}{Betzel, R.~F.}
	\newblock \bibinfo{title}{Modular {Brain} {Networks}}.
	\newblock \emph{\bibinfo{journal}{Annual Review of Psychology}}
	\textbf{\bibinfo{volume}{67}}, \bibinfo{pages}{613--640}
	(\bibinfo{year}{2016}).
	
	\bibitem{bassett_small-world_2017}
	\bibinfo{author}{Bassett, D.~S.} \& \bibinfo{author}{Bullmore, E.~T.}
	\newblock \bibinfo{title}{Small-{World} {Brain} {Networks} {Revisited}}.
	\newblock \emph{\bibinfo{journal}{The Neuroscientist}}
	\textbf{\bibinfo{volume}{23}}, \bibinfo{pages}{499--516}
	(\bibinfo{year}{2017}).
	
	\bibitem{tanner_functional_2023}
	\bibinfo{author}{Tanner, J.}, \bibinfo{author}{L., S.~M.},
	\bibinfo{author}{Coletta, L.}, \bibinfo{author}{Gozzi, A.} \&
	\bibinfo{author}{Betzel, R.~F.}
	\newblock \bibinfo{title}{Functional connectivity modules in recurrent neural
		networks: function, origin and dynamics}  (\bibinfo{year}{2023}).
	\newblock \urlprefix\url{https://arxiv.org/abs/2310.20601}.
	\newblock \bibinfo{note}{Publisher: arXiv Version Number: 1}.
	
	\bibitem{fries2009neuronal}
	\bibinfo{author}{Fries, P.}
	\newblock \bibinfo{title}{Neuronal gamma-band synchronization as a fundamental
		process in cortical computation}.
	\newblock \emph{\bibinfo{journal}{Annual review of neuroscience}}
	\textbf{\bibinfo{volume}{32}}, \bibinfo{pages}{209--224}
	(\bibinfo{year}{2009}).
	
	\bibitem{loebel2002computation}
	\bibinfo{author}{Loebel, A.} \& \bibinfo{author}{Tsodyks, M.}
	\newblock \bibinfo{title}{Computation by ensemble synchronization in recurrent
		networks with synaptic depression}.
	\newblock \emph{\bibinfo{journal}{Journal of computational neuroscience}}
	\textbf{\bibinfo{volume}{13}}, \bibinfo{pages}{111--124}
	(\bibinfo{year}{2002}).
	
	\bibitem{li2009consensus}
	\bibinfo{author}{Li, Z.}, \bibinfo{author}{Duan, Z.}, \bibinfo{author}{Chen,
		G.} \& \bibinfo{author}{Huang, L.}
	\newblock \bibinfo{title}{Consensus of multiagent systems and synchronization
		of complex networks: A unified viewpoint}.
	\newblock \emph{\bibinfo{journal}{IEEE Transactions on Circuits and Systems I:
			Regular Papers}} \textbf{\bibinfo{volume}{57}}, \bibinfo{pages}{213--224}
	(\bibinfo{year}{2009}).
	
	\bibitem{rulkov1995generalized}
	\bibinfo{author}{Rulkov, N.~F.}, \bibinfo{author}{Sushchik, M.~M.},
	\bibinfo{author}{Tsimring, L.~S.} \& \bibinfo{author}{Abarbanel, H.~D.}
	\newblock \bibinfo{title}{Generalized synchronization of chaos in directionally
		coupled chaotic systems}.
	\newblock \emph{\bibinfo{journal}{Physical Review E}}
	\textbf{\bibinfo{volume}{51}}, \bibinfo{pages}{980} (\bibinfo{year}{1995}).
	
	\bibitem{pecora1990synchronization}
	\bibinfo{author}{Pecora, L.~M.} \& \bibinfo{author}{Carroll, T.~L.}
	\newblock \bibinfo{title}{Synchronization in chaotic systems}.
	\newblock \emph{\bibinfo{journal}{Physical review letters}}
	\textbf{\bibinfo{volume}{64}}, \bibinfo{pages}{821} (\bibinfo{year}{1990}).
	
	\bibitem{lu2020invertible}
	\bibinfo{author}{Lu, Z.} \& \bibinfo{author}{Bassett, D.~S.}
	\newblock \bibinfo{title}{Invertible generalized synchronization: A putative
		mechanism for implicit learning in neural systems}.
	\newblock \emph{\bibinfo{journal}{Chaos: An Interdisciplinary Journal of
			Nonlinear Science}} \textbf{\bibinfo{volume}{30}} (\bibinfo{year}{2020}).
	
\end{thebibliography}
\end{document}